**Inner optimizers in active-learning Kriging reliability analysis: a pre-specified comparison and the unconservative estimates left by unlearned failure modes**[1]

Wei-Chang Yeh[1,*]

[1] Department of Industrial Engineering and Engineering Management, National Tsing Hua University, Hsinchu 300044, Taiwan

[*] Corresponding author: yeh@ieee.org

**Abstract**

Active-learning Kriging often replaces candidate pools by continuous metaheuristic search. A pre-specified comparison sharing surrogate, population and stopping rule isolated this inner optimizer: five budget-matched continuous optimizers, including hybrid swarm optimization with L-BFGS-B polishing, against candidate pools on eleven problems (dimension 2 to 100). The primary hypothesis was not supported: against optimizers sharing its criterion the hybrid won 6, tied 41 and lost 19 of 66 Holm-corrected accuracy comparisons, and against pool methods 0, 23 and 21 of 44. With sequential selection on three series systems, multistart L-BFGS-B left a failure mode unlearned in 59 of 60 runs, more often than swarm, genetic-algorithm or AK-MCS. In all 198 runs (22 post-freeze) with an unlearned mode, the learned modes did not compensate and failure probability was underestimated (exploratory). The error criterion can accept substantial underestimation: it accepted 17 runs (13 pool-selected) at 0.38 to 0.89 of the reference; 181 exhausted the budget. Follow-up analyses on single-design-point problems indicate underestimation when selections concentrate near the design point. Kriging-believer pool selection satisfied the surrogate stopping criterion with fewer calls than the best continuous method in all four informative batch cells (two post-freeze). We recommend default pool selection and coverage checks for any selection.

## 1. Introduction

**Context.** Active-learning Kriging (AK) replaces most evaluations of an expensive limit-state function $g$ by a Gaussian-process (GP) surrogate that is refined where it is most likely to misclassify the sign of $g$ [1–3]. Each refinement solves an inner optimization problem: find the point, or the batch of $q$ points, that maximizes a learning function. AK-MCS [1] solves it by enumeration over the Monte Carlo (MC) population that is also used to estimate the failure probability $P_f$. A growing line of work replaces enumeration by continuous search with gradient-based or population-based optimizers [4–8], and batch (parallel) variants choose $q$ points per round [9–15].

[1] Submitted to Reliability Engineering & System Safety. This is the full-length version of the manuscript; in the version submitted to the journal, some details and two tables are given as supplementary material.

**Gap.** When a new AK method is proposed, the learning function, the batch construction, the candidate space and the inner optimizer usually change at the same time, and the inner budget is rarely matched across methods. The contribution of the inner optimizer alone is therefore unknown. Related evidence exists in Bayesian optimization, where the maximization of acquisition functions has been studied separately [16, 17], and in our earlier controlled benchmark of SSO as an acquisition-function optimizer [18], but reliability analysis differs in two ways: the learning function is concentrated near the limit state, and a series system can have several separated failure regions (modes) that all contribute to $P_f$.

**Approach.** We isolate the inner optimizer. All methods share one GP model, one MC population, one initial design and one stopping rule per seed; continuous optimizers receive the same hard inner budget per round. The hypotheses, metrics, tests and multiplicity families were fixed before the main experiments in a prospectively frozen protocol (an internal, time-stamped record, not a public registry); later changes to the design were recorded as dated amendments, and every decision taken after the freeze is summarized in Section 5.1 (Table 3) and listed with its date in Appendix Table S1.

**Findings and contributions.**

1. A pre-specified negative main result. SSO with L-BFGS-B polishing (SSOLB) was not better than its opponents on accuracy or cost in 110 Holm-corrected comparisons per metric (Section 6.1).
2. A finding on failure modes. The inner optimizer changes which failure modes the surrogate learns: with sequential selection ($q = 1$), multistart L-BFGS-B left modes unlearned more often than plain SSO, GA-PQPVC and the pool method AKMCS-U (pre-specified secondary hypothesis H-MM, with a learned-mode criterion fixed after pre-tests of multistart L-BFGS-B and before the main experiments). In every mode-classified run with an unlearned mode, the learned modes did not compensate and $\hat{P}_f$ was below $P_f$ (exploratory; Section 6.2).

3. A practical default. The candidate-pool method AK-MCS was never significantly less accurate than the continuous optimizers it was compared with at $q = 1$ (exploratory), and in the pre-specified batch comparison, pool with Kriging-believer selection (POOL-KB-U) satisfied the surrogate-based stopping criterion with fewer limit-state calls than the best continuous method in all four informative cells, two of them on the four-mode series-system problem series4_d100 at $d = 100$ (a post-freeze addition; Sections 6.1 and 6.3).
4. A distinction between optimizer quality and estimator quality (exploratory; from the 23-bar truss, a post-freeze addition, and from follow-up analyses of existing runs). That direct maximization of a learning function can cluster the selected points is known [6, 19–21], and density-weighted learning functions are established [22–24]. We report a consequence not described there: on single-design-point problems, selections concentrated near the design point $u^*$ left the rest of the failure domain false-safe and $\hat{P}_f$ too low in the cases examined, and we propose that the density factor of the learning function contributes to this concentration, although it was not isolated experimentally (Sections 6.4 and 7.1).
5. A reproducible protocol: a prospectively frozen plan, amendment log, paired seeds, bit-identical re-computation of every per-mode estimate, and code and data release (Data and Code Availability).

**Scope of the claims.** The results hold for the tested problems ($d = 2$–$100$, analytical problems and two linear-elastic trusses), budgets and stopping rule. We do not claim that metaheuristics are never useful in AK; the plain-SSO results at $d = 100$ are mixed and exploratory (Section 6.3).

**Organization.** Section 2 lists acronyms, notation and assumptions. Section 3 reviews related work and states the problem. Section 4 describes the AK framework and the compared methods, Section 5 the pre-specified design, Section 6 the results, Section 7 the discussion with a selection table, and Section 8 concludes.

## 2. Acronyms, Notations, Nomenclatures, and Assumptions

This section collects the abbreviations, symbols, terms with a specific meaning and the assumptions used throughout.

**Acronyms.** AK, active-learning Kriging; AK-MCS, active-learning Kriging with Monte Carlo simulation [1]; AKMCS-U, the AK-MCS pool method with the $U$ function ($q = 1$);

CMA-ES, covariance matrix adaptation evolution strategy [25]; CONT-, prefix of a continuous-search method; E1, relative error of $\hat{P}_f$ at stop; E2, number of limit-state calls at stop; E4, success rate; GA, genetic algorithm; GP, Gaussian process; H-MM, pre-specified secondary hypothesis on missed modes; HL, Hodges–Lehmann estimator; IF, influence function; KB, Kriging believer [26]; L-BFGS-B, limited-memory BFGS with bounds [27]; MC, Monte Carlo; PAKP, parallelized adaptive Kriging with partial least squares [12]; POOL-, prefix of a candidate-pool method; PQPVC, the pseudo-posterior learning function of [8]; SSO, simplified swarm optimization [28, 29]; SSOLB, SSO followed by L-BFGS-B polishing; TOST, two one-sided tests; W/T/L, win/tie/loss.

**Notations.** Table 1 lists the symbols.

Table 1. Notation.

| Symbol | Meaning |
|---|---|
| $u$, $d$ | point in standard normal space; dimension |
| $g(u)$ | limit-state function; failure if $g(u) \leq 0$ |
| $g_k(u)$ | component $k$ of a series system, $g = \min_k g_k$ |
| $P_f$, $\hat{P}_f$ | reference and estimated failure probability |
| $\hat{N}_f$ | number of points of $S$ predicted to fail, $\#\{u \in S: \mu(u) < 0\}$ |
| $P_k$, $\hat{P}_k$ | reference and estimated probability of failure governed by mode $k$ |
| $\beta$ | reliability index, $\beta = -\Phi^{-1}(P_f)$; $\Delta\beta = \hat{\beta} - \beta_{\mathrm{ref}}$ |
| $\mu(u)$, $\sigma(u)$ | GP posterior mean and standard deviation |
| $U(u)$ | AK-MCS learning function, $U = \lvert\mu\rvert/\sigma$ [1] |
| $S$, $N_{\mathrm{MC}}$ | common MC population of one seed and its size |
| $n_0$ | size of the initial design |
| $q$ | batch size (points per round) |
| $B$ | inner budget per round (learning-function evaluations of one q-point batch) |
| $R$ | half-width of the continuous search box $[-R, R]^d$ |
| $\hat{\varepsilon}$, $\varepsilon_{\mathrm{stop}}$ | estimated relative error used for stopping; threshold 0.02 |
| $N_{\mathrm{call}}$, $N_{\max}$ | limit-state calls at stop; call cap $100 + 10d$ |
| E1, E2 | $\lvert\hat{P}_f - P_f\rvert/P_f$ at stop; $N_{\mathrm{call}}$ at stop |
| E4 | success rate: share of runs that reached E1 ≤ 0.05 and did not stop at $N_{\max}$ |

**Nomenclatures.**

- Inner budget. One unit of $B$ is one evaluation of the learning function for one q-point batch, with or without its gradient. A GA step that scores one point counts 1/q unit (GA receives $q \cdot B$ single-point evaluations). Pool methods enumerate $S$ and are not budget-matched (Section 4.2).
- Mode covered (Amendment A3). At least one evaluated point lies within first-order distance $\tau = 1.0$ of the component limit surface $g_k = 0$, with $k$ the nearest component.
- Mode learned (Amendment A6 (b)). At stop, $0.5 \leq \hat{P}_k / P_k \leq 2$.
- Mode-wise probability. Each point $u$ of $S$ with $\mu(u) < 0$ is assigned to its governing component $k(u) = \text{argmin}_k g_k(u)$ of the true series system (ties have probability zero; the lowest $k$ is taken), and $\hat{P}_k = \#\{u \in S: \mu(u) < 0, k(u) = k\}/N_{\text{MC}}$. Overlapping failure regions are thus split by the governing component, and $\sum_k \hat{P}_k = \hat{P}_f$; a point that is predicted to fail but is safe is counted in the mode of its governing component. The reference $P_k$ is the same split of a $10^9$-sample MC population with $g(u) \leq 0$ in place of $\mu(u) < 0$, so the $P_k$ sum to the MC estimate of $P_f$ (values in Section 5.2).
- Unconservative. $\hat{P}_f < P_f$, i.e. $\Delta\beta > 0$: reliability is overestimated.
- Exploratory. Any analysis not specified before the data were seen.

**Assumptions.**

A1. All inputs are mapped to independent standard normal variables $u$; the mapping is exact for the tested problems (independent normal, lognormal and Gumbel marginals).

A2. The reference $P_f$ is exact or estimated with a coefficient of variation small enough not to affect E1 (about 0.10% or less for all problems used; Section 5.2).

A3. Limit-state calls dominate the cost of the analysis; $N_{\text{call}}$ is the cost measure. Inner-loop time is reported separately and is not used for the main comparisons (Section 5.4).

A4. One GP model class with fixed hyperparameter-refit schedule is used for all methods; differences between methods are attributed to the selection step only.

## 3. Related Work and the Proposed Problem

### 3.1. Related Work

**Learning functions and pool-based AK.** The expected feasibility function [2], the $U$ function of AK-MCS [1] and stepwise uncertainty reduction [3, 30] define where the surrogate is refined. Pool-based variants restrict the candidates to an MC, importance-sampling or subset-simulation population [1, 31–34], and stopping rules based on the estimated error of $\hat{P}_f$ were proposed in [35, 36]. Reviews and benchmarks [37, 38] compare learning functions, surrogates and stopping criteria, but keep the candidate pool fixed.

**Continuous search and metaheuristics in the inner loop.** Several methods search the learning function continuously: radial-basis surrogates with GA [4], PSO in a hollow hypersphere [5], PSO with a modified $U$ [6, 7], and the quasi-Bayesian active learning of [8], which maximizes a pseudo-posterior learning function with a GA in $[-R, R]^d$. Sample-pool-free approaches extend this to high dimension [39]. In these studies the inner optimizer is introduced together with a new learning function or batch rule, and inner budgets are not matched against a pool baseline.

**Batch selection.** Kriging believer [26], k-means clustering of candidates [10, 40], local penalization [41], influence-function penalties [11, 12, 42] and multi-point Bayesian criteria [13–15, 43, 44] select $q$ points per round. Among the studies whose full text we could verify, the largest dimension in which continuous batch search was reported was $d \leq 15$ [8, 9, 13–15]; the batch methods applied at $d \geq 20$ used candidate pools [12, 45].

**Multiple failure modes.** Series systems with several design points are a known difficulty for AK [46–49]; the four-branch function [1, 50] is the common test. In the studies we reviewed, missed modes are reported as a qualitative failure, not measured per mode.

**Clustering of selected points.** That direct maximization of a learning function can place new points in clusters is known. In a pool-free PSO method, Li et al. [6] note that the selected samples "may display a concentrated distribution in certain areas" and that an effective optimizer can find a near-duplicate optimum whose benefit to the Kriging model is limited; they add a distance-based density control. Zhang et al. [19] name clustered or overlapping training samples as a drawback of arg-max updating; distance constraints or penalty terms against clustered or redundant samples were proposed in [20, 21, 51], and clustering of candidates (k-means, DBSCAN) [9, 10, 40, 52, 53], penalization [11, 12, 41, 42] and the Kriging believer [26, 54] serve the same purpose in batch selection. In these works the stated cost of clustering is wasted limit-state evaluations; none of them, as far as we found, reports a bias of $\hat{P}_f$. Our contribution is limited to that consequence, its proposed density-weighting mechanism and the follow-up evidence (Section 1). Bect et al. [3] regard restricting the search to the MC sample as a simplification that "is obviously sub-optimal", and the benchmark of [38] likewise calls the discrete approximation "not an optimal approach". In contour estimation, Cole et al. [55] observe that some stochasticity in the inner search helps to avoid pathologies. Weighting the learning function by the probability density is also established. AWL-MCS [22] weights its learning function by the joint PDF and minimizes it with PSO. PAK-Bn [24] multiplies the U-type criterion by a penalty that discourages candidates far from the estimated nearest failure point while that estimate is still changing; its authors show that the selected points lie mostly in the critical regions near the design points and note that regions where the surrogate remains inaccurate "have little impact on the failure probability estimation due to the low probability density". According to its abstract, AK-PDF [23] gives candidates in regions of higher probability density more weight to be selected; we could not access its full text and rely on the abstract only. The two studies we read in full report accurate estimates on their examples and do not report a bias of $\hat{P}_f$. In the literature we reviewed, we found no report that concentration of the selected points near the design point biases $\hat{P}_f$, or that a better maximizer of the learning function gives a worse estimate. For the problems tested here, our results qualify the argument of [24]: on the 23-bar truss the mass left false-safe away from $u^*$ was not negligible and accounted for almost all of the underestimation (a post hoc observation on a post-freeze addition, examined by the follow-up analyses of Section 6.4).

**Inner optimizers in Bayesian optimization.** Studies of acquisition-function maximization [16, 17] and our controlled benchmark of SSO as an acquisition-function optimizer [18] show that the inner optimizer can change both the solution quality and the inner-loop time; the median inner-time ratio of SSO with L-BFGS-B polishing to multistart L-BFGS-B was 0.31× over the 14 qLogEHVI settings with timing data (range 0.08–0.57) [18, Table 6].

**Gap.** In the literature we reviewed, no study isolated the inner optimizer in AK reliability analysis under equal inner budgets, with paired seeds and a common pool baseline, or measured how the optimizer affected the per-mode accuracy of $\hat{P}_f$ in series systems. Nor did we find the known clustering of maximized learning functions linked to a bias in $\hat{P}_f$.

### 3.2. Proposed Problem

**Problem P1: effect of the inner optimizer.** For a limit state $g$ with $P_f = P(g(u) \leq 0)$, $u \sim N(0, I_d)$, and an AK loop that differs only in the inner optimizer $o \in O$, estimate the paired differences

$$\Delta_{\mathrm{E1}}(o, o') = \mathrm{E1}(o) - \mathrm{E1}(o'), \quad \Delta_{\mathrm{E2}}(o, o') = \mathrm{E2}(o) - \mathrm{E2}(o'), \tag{1}$$

over seeds that fix $S$, the initial design and the GP settings, for every problem, batch size $q \in \{1,4,8\}$ and opponent $o'$. The primary method of the frozen protocol is $o$ = SSOLB. Gap closed: no study we found isolates the inner optimizer under equal inner budgets (Section 3.1). Budget unit: $B$ learning-function evaluations of a q-point batch per round (Section 2); the outer cost unit is the limit-state call.

**Problem P2: mode-wise accuracy.** For a series system $g = \min_k g_k$ with separated design points, estimate for each method the probability that at stop some mode is not learned,

$$\pi(o) = P\big(\exists k: \hat{P}_k / P_k \notin [0.5, 2]\big), \tag{2}$$

and the consequence for $\hat{P}_f / P_f$ and $\Delta\beta$. Gap closed: missed modes have been reported qualitatively [46–49] but not measured per mode against the inner optimizer. Budget unit: as for Problem P1.

**Problem P3: point concentration and estimator bias (post-freeze addition).** For a single-design-point limit state with FORM design point $u^*$, a reconstructed AK state $s$ and a GP mean $\mu_s$, let $\mathrm{FS}(s) = P(g(u) \leq 0, \mu_s(u) \geq 0)$ be the false-safe probability. Adding one point chosen by rule $k \in \{A, B\}$ (A: the best-found reference maximizer of the criterion of Equation (4), Section 6.4; B: the AK-MCS pool choice) and refitting gives $\mathrm{FS}_k(s)$, and the problem is to test whether

$$\delta(s) = \mathrm{FS}_A(s) - \mathrm{FS}_B(s) > 0 \tag{3}$$

in most states, and whether, across methods and problems, the distance of the selected points from $u^*$ is positively rank-correlated with $\hat{P}_f / P_f$. Gap closed: clustering of selected points is known [6, 19], but not as a source of bias in $\hat{P}_f$ (Section 3.1). Budget unit: one added limit-state call per state; the analyses use saved states and selections only (Section 6.4).

## 4. Compared Methods

### 4.1. Active-Learning Kriging Framework and Learning Functions

**Loop.** Figure 1 shows the AK loop used for every method. The MC population, the initial design, the GP model and the stopping rule are shared per seed; the compared methods differ only in the selection step, which either enumerates the population or searches the continuous space with a hard inner budget.

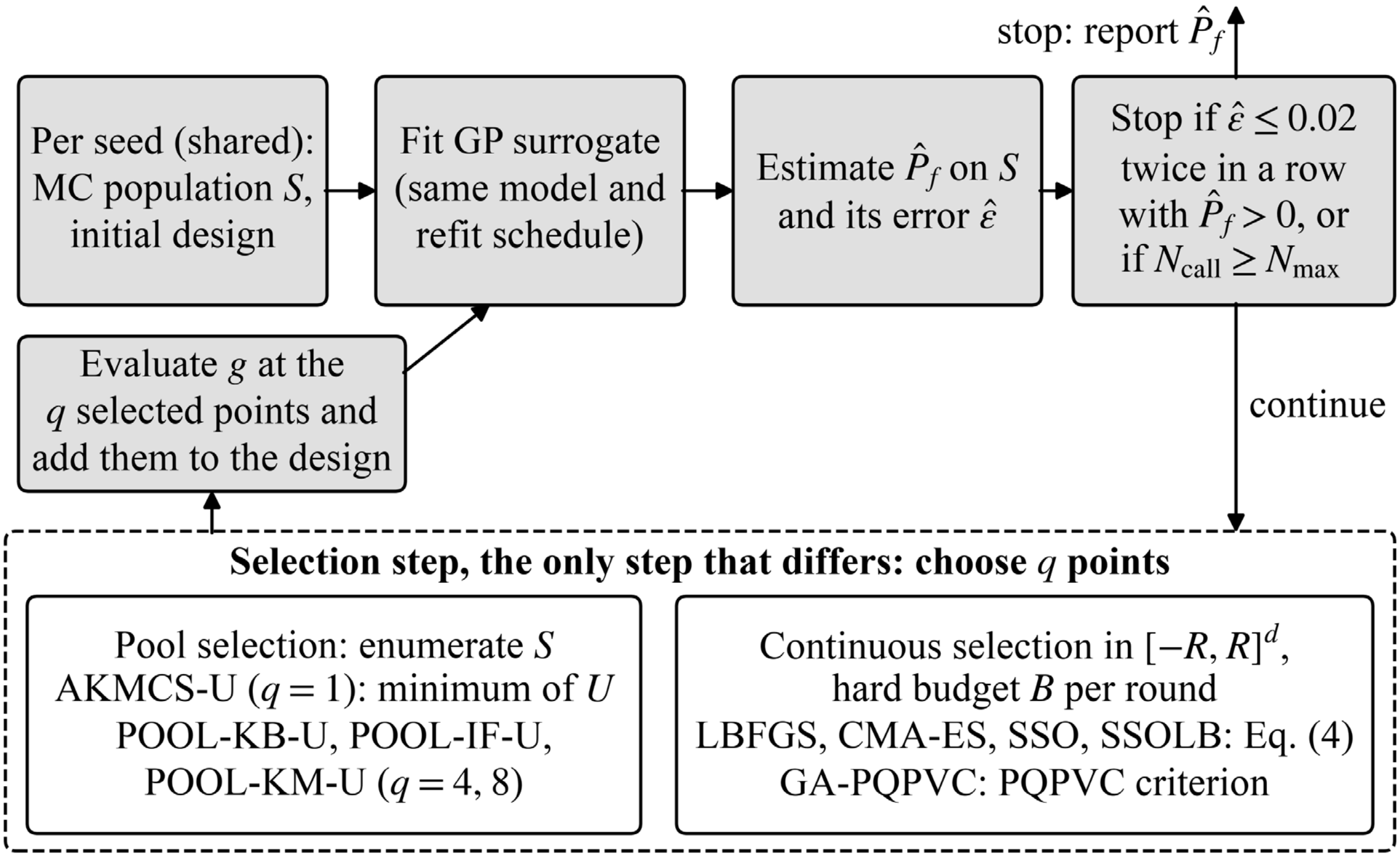

Figure 1. The active-learning Kriging loop of this study. Shaded: steps shared by all methods of one seed (MC population $S$, initial design, GP model, estimate of $P_f$ and stopping rule). Unshaded: the selection step, the only step in which the methods differ; pool methods enumerate $S$, continuous methods search the box $[-R, R]^d$ with a hard budget of $B$ learning-function evaluations per round. $q$: batch size; $\hat{\varepsilon}$: estimated relative error of $\hat{P}_f$; $N_{\max}$: call cap.

**Surrogate.** A GP with constant mean and ARD squared-exponential kernel (BoTorch SingleTaskGP [56]), inputs scaled from $[-R, R]^d$, standardized outputs, float64. Hyperparameters are refitted every round for $d \le 20$ (all problems except $d = 100$) and every five rounds with warm start for $d = 100$, with a final full refit at stop; the schedule is identical for all methods. Predictions on $S$ use an exact cached-Cholesky predictor validated against GPyTorch to a relative difference $\le 2.6 \times 10^{-10}$.

**Population and estimate.** For each seed one MC population $S$ of size $N_{\text{MC}} = \max(10^6, \lceil 2500/P_f \rceil)$ is drawn and shared by all methods; $\hat{P}_f = \#\{u \in S: \mu(u) < 0\}/N_{\text{MC}}$. For the original 56-bar truss threshold ($P_f \approx 4.9 \times 10^{-8}$, $d = 11$), $S$ is replaced for each seed by an importance-sampling population of $N = 10^6$ points drawn from $h = N(u^*, I)$, a single Gaussian with identity covariance centred at the FORM design point $u^*$ ($\beta = 5.36$), shared by all methods of the seed. $\hat{P}_f$ and $\hat{\varepsilon}$ then use the likelihood-ratio weights $w(u) =$

$\varphi_d(u)/h(u)$: $\hat{P}_f = \sum_{u\in S} w(u)\,\mathbf{1}[\mu(u) < 0]/N$, and in $\hat{\varepsilon}$ (below) each term of the sum is multiplied by $w(u)$ and $\max(1, \hat{N}_f)$ is replaced by $\sum_{u\in S} w(u)\,\mathbf{1}[\mu(u) < 0]$ ($\hat{\varepsilon}$ is infinite if no point of $S$ is predicted to fail). No selection criterion uses the weights $w(u)$: the pool methods apply their rules based on the unweighted $U$ (Section 4.2) to these importance-sampling candidates, which lie around $u^*$, whereas the continuous methods maximize their $\varphi_d$-weighted criteria (Equation (4); PQPVC for CONT-GA-PQPVC) over the box, as on every other problem; the initial design is drawn from this population as well. The population has its own sampling uncertainty: with exact classification (the limit state evaluated at every point), $\hat{P}_f/P_f$ over 20 independent populations of this design had a standard deviation of 0.0024 (divisor 20; 0.0025 with divisor 19), i.e. a coefficient of variation of $\hat{P}_f$ of 0.24% (0.25% with divisor 19), and an E1 of 0.0020 on average and at most 0.0048 (post hoc; new CPU draws, not the runs' own populations). E1 values of this order on this problem are within the sampling uncertainty of the estimator.

**Initial design and search space.** $n_0 = \max(12, d + 1)$ points drawn from $S$, identical for all methods. Continuous search uses $[-R, R]^d$ with $R = \sqrt{(\chi_d^2)^{-1}(1 - 10^{-10})}$ as in [8, 14].

**Stopping.** $\hat{\varepsilon} = \sum_{u\in S} \Phi(-|\mu(u)|/\sigma(u))/\max(1, \hat{N}_f) \le 0.02$ in two consecutive rounds with $\hat{P}_f > 0$ (Amendment A1), or $N_{\text{call}} \ge N_{\max} = 100 + 10d$, where $\hat{N}_f = \#\{u \in S: \mu(u) < 0\}$ is the number of points of $S$ predicted to fail.

**Pool learning function.** Pool methods minimize $U = |\mu|/\sigma$ over $S$ [1].

**Continuous joint batch criterion.** Minimizing $U$ in continuous space is degenerate ($U \to 0$ anywhere on the predicted limit state, including far tails), so continuous methods maximize a density-weighted joint criterion over the $q \times d$ batch $X = (x_1, \dots, x_q)$:

$$\text{logqLF}(X) = \sum_i \left[\log\Phi\left(-\frac{|\mu(x_i)|}{\sigma(x_i)}\right) + \log\varphi_d(x_i)\right] + \sum_{i<j} \log\left(1 - \rho_{ij}^2\right), \quad (4)$$

where $\rho_{ij}$ is the GP posterior correlation of $x_i$ and $x_j$; the last term repels points of one batch. The criterion of Equation (4) is defined in this study; it is related to the influence-function and penalty schemes of [11, 12, 41, 42]. The negative control (Block C, exploratory) is consistent with the degeneracy of continuous raw-U search (Section 6.3).

**GA with PQPVC.** CONT-GA-PQPVC approximately reproduces [8]: the pseudo-posterior learning function of [8, Eqs. (35)–(36)] is maximized by a GA, $q$ times sequentially.

### 4.2. Inner Optimizers and Candidate-Pool Baselines

**Continuous inner optimizers.** CONT-LBFGS, CONT-CMAES, CONT-SSO and SSOLB act on the same criterion, Equation (4); CONT-GA-PQPVC maximizes the PQPVC criterion of [8] (Section 4.1), so comparisons with it change the criterion as well as the optimizer. All five stop when the hard budget $B$ is reached; a meter wrapped around the criterion stops the optimizer before the next evaluation would exceed $B$. $B$ was calibrated per (problem, $q$) as the median per-round use of unbudgeted multistart L-BFGS-B on pilot seeds 1000–1004 (Table 2).

- CONT-LBFGS: multistart L-BFGS-B [27] via BoTorch optimize_acqf, 10 restarts from 512 raw samples, scipy maxfun $\lfloor (B - 512)/10 \rfloor$. Each evaluation includes the gradient (autograd) and counts as one unit.
- CONT-CMAES: CMA-ES [25], pycma 4.5.0 with increasing-population restarts (IPOP [57]), $\sigma_0 = 0.2$ in the search coordinates scaled to [0, 1] ($0.4R$ in $u$-space; 3.30 for the 23-bar truss, a post-freeze addition), 10% of $B$ for a random warm start.
- CONT-SSO: plain SSO [28, 29], 100 particles, each a flattened $q \times d$ vector, synchronous global best, Cg/Cp/Cw = 0.65/0.90/0.95, uniform initialization in $[-R, R]^{qd}$, no gradient.
- CONT-SSOLB: SSO with 75% of $B$, then L-BFGS-B from the five best distinct personal bests with the remaining 25%.
- CONT-GA-PQPVC: population $\min(200, 10qd)$; $q \cdot B$ single-point evaluations per round ($B$ point-equivalents). The SSO, SSOLB and CMA-ES implementations are the unmodified code of [18, 58].

Table 2. Hard inner budget $B$ (learning-function evaluations per selection round) per (problem, $q$).

| Problem | $d$ | $q = 1$ | $q = 4$ | $q = 8$ |
|---|---|---|---|---|
| four-branch | 2 | 3367 | 7997 | 14717 |
| fourbranch_k6 | 2 | 3072 | 6812 | 13517 |
| oscillator | 6 | 3712 | 10272 | 17342 |
| Rackwitz $d = 2$ | 2 | 3967 | 6732 | 8047 |
| Rackwitz $d = 6$ | 6 | 3742 | 11282 | 15522 |
| Rackwitz $d = 10$ | 10 | 3462 | 11352 | 14827 |
| Rackwitz $d = 100$ | 100 | 2672 | 7142 | 9162 |

| Problem | $d$ | $q = 1$ | $q = 4$ | $q = 8$ |
|---|---|---|---|---|
| 56-bar truss | 11 | 2832 | 6487 | 11912 |
| 56-bar truss, mod. | 11 | 3062 | 7462 | 12372 |
| series4_d20 | 20 | 2597 | 5892 | 10852 |
| series4_d100 | 100 | 2652 | 5007 | 8762 |
| 23-bar truss | 10 | 3512 | 10692 | 15727 |

$B$ is the median per-round use of unbudgeted multistart L-BFGS-B on pilot seeds 1000–1004. Rackwitz $d = 100$ at $q = 1$ was calibrated but not run (no Block B arm at $d = 100$ except series4_d100). The 23-bar truss values were calibrated for Stage A11 (post-freeze addition); series4_d100 is also a post-freeze addition (Section 5.1).

**Candidate-pool baselines.** Four pool methods select from $S$.

- AKMCS-U ($q = 1$): $\text{argmin}_{u \in S}\, U$ [1].
- POOL-KB-U ($q = 4$, 8): Kriging believer [26, 54] with fixed hyperparameters and an exact rank-1 variance update (verified against conditioning on observations, relative difference < $10^{-6}$).
- POOL-IF-U ($q = 4$, 8): first point $\text{argmin}_S\, U$ as AK-MCS; points 2…q by the multiplicative influence-function penalty of PAKP [12, Eqs. (22)–(24)], without its PLS reduction.
- POOL-KM-U ($q = 4$, 8): weighted k-means on candidates with $U < 2$ [10, 40] (not an H0-main opponent; included in H-MM and in the pool set of H2 at $d = 20$). Pool methods score all of $S$ each round and are not budget-matched; their evaluation counts and times are reported as they are (Section 5.4). POOL-LP-U [41] and CONT-PSO-R22 [6] were pre-specified but not run: the full texts needed to verify their definitions were not available before the experiments ended, and both were removed from the design (Amendments A7 and A13). Neither was an H0-main opponent or an H-MM comparison, and no hypothesis definition changed; under the frozen plan, POOL-LP-U would have been one of the pool methods from which the best pool method of H2 is chosen.

## 5. Experimental Design

### 5.1. Prospectively Frozen Protocol and Amendments

The protocol was frozen internally before the main experiments; it was not deposited in a public registry. The frozen plan file calls itself a pre-registration, a term this paper avoids for that reason. The record consists of the frozen plan file, which fixes the hypotheses, methods, budgets, stopping rule, metrics, tests and families and records the SHA-256 of the plan text and of its addendum on the 56-bar truss and the code version at the freeze; the SHA-256 of the frozen plan file itself (32ba54a11b6e1b4dd20503b0938ebe9cdff7d31d9eeeb07d5798cc30799dcc4d); its commit to the author's version-control repository on 2026-10-11 at 13:41 UTC (commit 4d9254d); and the dated amendment files. The frozen plan file and the amendment files are in the data package (Data and Code Availability); the plans of the Stage A11 diagnosis and of the follow-up analyses of Section 6.4, the Stage A11 compute-cap decisions, the budget-calibration rules, the commit log of the repository and the job records of the NCHC Nano4 cluster will be added to it before publication. The start times recorded by the runs on the cluster are consistent with this order: the first main-experiment run (Stage S1) started on 2026-10-12 at 22:31 UTC, about 33 hours after the freeze commit, and Stages A10 and A11 started on 2026-10-14 and 2026-10-15.

After the freeze the design was changed only by 13 dated amendments, each approved before it took effect (Amendments A1–A13; Appendix Table S1 states for each whether it was written before or after the affected results were seen). Amendments A1–A8 were made before Stage S1 and Amendments A9–A13 after it; among the latter, Amendment A10 replaced the planned stages S1b and S2 by two $d = 100$ problems (rackwitz_d100, which is in the frozen plan, and series4_d100, a post-freeze addition), Amendment A11 added the 23-bar truss (a post-freeze addition), Amendment A12 reduced the H3 timing replay and Amendment A13 removed two methods of the frozen plan that had not been run.

Table 3 summarizes, and Appendix Table S1 lists with record, date, reason and status, the amendments, the additions made before any comparison output and every addition, procedure and analysis fixed after results had been seen. The additions made before any comparison output existed (Amendments A3, A4 and A6, all before Stage S1) are treated as part of the pre-specified design. Post-freeze additions are the four new parts added after results had been seen: series4_d100 (Amendment A10), the 23-bar truss case (Stage A11, Amendment A11) and the two follow-up analyses of the concentration mechanism, specified with predictions and pass thresholds after the initial observation and before their outputs were generated (Section 6.4); they are labelled as such at their first mention in each section and in each table and figure caption that uses them. Analyses of existing data added after results had been seen (the diagnosis and the sensitivity analysis of Section 6.4, the robustness checks of H2 and H3 and the re-evaluation on an independent MC population in Section 6.1, the compensation analysis of Section 6.2, the direction audit of Section 5.4 and the other analyses in Appendix Table S1) are labelled exploratory or post hoc where they are reported. The tests on series4_d100 are the pre-specified ones, in the pre-specified families, but the problem was chosen after S1 and because of its result; it is the only problem of Stage A10 that is not in the frozen plan, so results reported for Stage A10 as a whole include it. The truss case and the follow-up analyses are outside the original families. We call an analysis confirmatory only when it tests a prediction on data generated after the prediction (the pre-specified tests of S1 and Stage A10); analyses of existing runs are follow-up analyses, even when they were specified before their outputs were generated. Procedures fixed after results had been seen (Amendments A9 and A12, the budget-calibration rules, the analysis-time choices for H1 and H2, the restriction of their results to the Block A cells, and the Stage A11 compute caps) are disclosed where they apply, and analyses introduced after results had been seen are interpreted accordingly.

Table 3. Summary of the changes to the frozen design and of the analyses fixed after results had been seen (freeze: 2026-10-11, 13:41 UTC). Every item, with its record, date and time, reason and status: Appendix Table S1.

| Category | Items (record dates, UTC) | Status in this paper |
|---|---|---|
| Additions before any comparison output | H-MM (Amendment A3, 2026-10-12); fourbranch_k6 and series4_d20 (Amendment A4, 2026-10-12); learned-mode criterion and sensitivity analyses (Amendment A6, 2026-10-12, after S0, before S1) | pre-specified |
| Routine amendments | Amendments A1, A2, A5, A7 and A8 (2026-10-11 to 2026-10-12, before S1); Amendment A13 (2026-10-16: two methods that had not been run removed) | routine amendment |
| Post-freeze additions | series4_d100 (Amendment A10, 2026-10-14, after S1); 23-bar truss, Stage A11 (Amendment A11, 2026-10-15); two follow-up analyses of the concentration mechanism (2026-10-15, before their outputs) | post-freeze addition; the truss case and the follow-up analyses outside the original families |

| Category | Items (record dates, UTC) | Status in this paper |
| --- | --- | --- |
| Procedures fixed after results had been seen | TOST procedure (Amendment A9, 2026-10-14); budget-calibration rules (2026-10-14 and 2026-10-15); Stage A11 compute caps (2026-10-15); operational details of H1 and H2 (2026-10-16, before their output); H3 replay scope (Amendment A12, 2026-10-16); H1 and H2 restricted to the Block A cells (2026-10-17, after their output) | procedure or analysis choice; S1 TOST and the $q = 1$ cells of H1 and H2 exploratory |
| Analyses of existing runs after results had been seen | Stage A11 diagnosis and sensitivity analysis (2026-10-15); H2 fixed-pair check (2026-10-19); H0-main split and direction audit, H3 seed-level check, re-evaluation on an independent MC population (2026-10-21); compensation, E2 censoring, H2 test details, single-cell Holm p (2026-10-21); stop reasons, H0-main without truss56_orig, seed-level intervention check, importance-sampling uncertainty (2026-10-22); other analyses not named in the frozen plan | post hoc, exploratory, labelled where reported |

Dates are those of the commits to the author's repository (UTC). S0: pre-tests, inner-budget calibration and reproducibility checks (Section 5.3).

Amendment A11 was approved by the author and became effective on 2026-10-15; its runs form Stage A11. Its compute cap was raised three times, from 8 to 9 GPU-hours after the pilot runs and then to 11 and to 11.2 GPU-hours during the main runs, each time by a recorded decision of the author after the pre-specified re-projection exceeded the current cap. The total use was 10.92 GPU-hours: 2.92 GPU-hours over the original cap of 8 and 1.92 over 9, and within the caps of 11 and 11.2. Methods, seeds, budgets, stopping and metrics were not changed.

### 5.2. Test Problems

Table 4 lists the test problems with their reference failure probabilities.

Table 4. Test problems (reference values of this study; literature values as cross-checks).

| Problem | $d$ | Modes | Reference $P_f$ (this study) | Definition source |
|---|---|---|---|---|
| four-branch, $k = 7$ | 2 | 4 | $2.2249 \times 10^{-3}$ ($10^9$ MCS) | [1, Eq. (17)] |
| four-branch, $k = 6$ (fourbranch_k6) | 2 | 4 | $4.4536 \times 10^{-3}$ ($10^9$ MCS) | [1, 50] |
| oscillator (normal version) | 6 | 1 | $2.8596 \times 10^{-2}$ ($10^9$ MCS) | [1, 50, 59]; original lognormal version [60] not used |
| Rackwitz sum of lognormals | 2, 6, 10, 100 | 1 | 4.9214, 3.2087, 2.7252, 1.7338 ($\times 10^{-3}$) | [1, 61] |
| 56-bar space truss, $\Delta = 50$ mm | 11 | 1 | $4.873 \times 10^{-8}$ (IS, COV 0.058%) | probability model [8]; geometry [62] |
| 56-bar space truss, modified $\Delta = 28.1$ mm | 11 | 1 | $9.7594 \times 10^{-4}$ ($10^9$ MCS) | as above, threshold modified |
| series4_d20 | 20 | 4 | $3.5073 \times 10^{-3}$ (analytical) | defined in this study (Amendment A4) |
| series4_d100 | 100 | 4 | $3.5073 \times 10^{-3}$ (analytical) | defined in this study (Amendment A10; post-freeze addition) |

For the 56-bar truss, $E = 206\,\text{GPa}$ ([8] prints 2.06 GPa, which would give $P_f \approx 1$); the analysis is linear-elastic with small displacements, so given $E$ and $A$ the deflection is linear in the loads. Our reference for the original threshold is 1.35% below the value of [8, Table 7].

Series-system benchmarks. The four multi-mode problems are series systems $g = \min_k g_k$ with four components. Four-branch [1, 50]: $g_1 = 3 + 0.1(u_1 - u_2)^2 - (u_1 + u_2)/\sqrt{2}$, $g_2 = 3 + 0.1(u_1 - u_2)^2 + (u_1 + u_2)/\sqrt{2}$, $g_3 = (u_1 - u_2) + k/\sqrt{2}$ and $g_4 = (u_2 - u_1) + k/\sqrt{2}$, with $k = 7$ (four-branch) or $k = 6$ (fourbranch_k6). series4_d20 and series4_d100 (defined in this study; Amendments A4 and A10): $g_1 = 3 - u_1$, $g_2 = 3 + u_1$, $g_3 = 3.25 - u_2$ and $g_4 = 3.5 + u_2$ in $d = 20$ or $d = 100$ variables, of which $u_3, \dots, u_d$ do not enter $g$, so that $P_f = 1 - [1 - 2\Phi(-3)][1 - \Phi(-3.25) - \Phi(-3.5)] = 3.5073 \times 10^{-3}$. Reference mode-wise probabilities $P_1, \dots, P_4$ ($10^9$-sample MC split by the governing component; Section 2): four-branch 8.78, 8.80, 2.33 and 2.32, fourbranch_k6 8.79, 8.78, 13.49 and 13.51, and series4 13.50, 13.48, 5.76 and 2.32 (all $\times 10^{-4}$).

Engineering demonstration (Stage A11, post-freeze addition, exploratory). The 23-bar planar truss introduced by Lee and Kwak [63] ($d = 10$; lognormal moduli $E_c$, $E_d$ and areas $A_c$, $A_d$; Gumbel loads P1–P6; chord members share $E_c$ and $A_c$, diagonals share $E_d$ and $A_d$; these are the variables numbered 1 and 2 in [63]) and used in later surrogate studies [64–67]. The original limit state is $g = 0.11\,\mathrm{m} - V_1$, with $V_1$ the midspan deflection [63]; we use the 12 cm threshold of [66, 67] ($g = 0.12\,\mathrm{m} - V_1$, $P_f \approx 1.5 \times 10^{-3}$). Node coordinates are not tabulated in any source and were taken from the figure of [63] (six 4 m bays, height 2 m). The original analysis used a commercial finite-element code without stating the analysis type [63]; we use a linear-elastic, small-displacement direct-stiffness model. With $10^7$ samples it reproduces the published mean, standard deviation, skewness and kurtosis of $V_1$ (0.0794 m, 0.0111 m, 0.493 and 3.45 vs 0.0794 m, 0.0111 m, 0.492 and 3.46 [64]) and the published Monte Carlo $P_f$ ($8.79 \times 10^{-3}$ vs $8.70 \times 10^{-3}$ at 11 cm [65]; $1.50 \times 10^{-3}$ vs $1.52 \times 10^{-3}$ at 12 cm [66]). Our reference value from $10^9$ Monte Carlo samples is $P_{\mathrm{ref}} = 1.5177 \times 10^{-3}$ (coefficient of variation 0.08%, $\beta = 2.964$). All loads increase the midspan deflection, and FORM from 20 starting points (the origin and 19 random starts from $N(0, 2^2 I)$) converged to the same point ($\beta_{\mathrm{FORM}} = 3.172$); we take this as evidence that the limit state has a single design point, not as proof that the design point is unique. The truss therefore demonstrates accuracy only (all its runs stopped at the call cap; Section 6.4) and cannot test H-MM; the safety evidence in this paper comes from the multi-mode benchmarks (four-branch and series systems). A 3-bay, 5-storey frame ($d = 21$) was considered and deferred because its truncation convention could not yet be verified.

### 5.3. Blocks, Stages, Metrics and Hypotheses

**Blocks and stages.** The runs are organized in four blocks and three main stages, preceded by Stage S0 (pre-tests, inner-budget calibration and reproducibility checks on the computing platform; no S0 output enters a comparison).

- Block A (confirmatory): $q \in \{4,8\}$; POOL-KB-U, POOL-KM-U, POOL-IF-U, CONT-GA-PQPVC, CONT-LBFGS, CONT-CMAES, CONT-SSO, CONT-SSOLB; seeds 0–19.
- Block B ($q = 1$ reference): AKMCS-U, CONT-LBFGS, CONT-SSOLB, CONT-GA-PQPVC, CONT-CMAES, CONT-SSO ($d \leq 20$); at $d = 100$ only series4_d100 (a post-freeze addition) with AKMCS-U, CONT-LBFGS, CONT-SSOLB, CONT-GA-PQPVC, CONT-SSO.
- Block C (negative control, exploratory): {pool, CONT-LBFGS, SSOLB} × {raw $U$, density-weighted} (six arms), four-branch, $q = 4$, 5 seeds.
- Block D (budget sensitivity, exploratory): Rackwitz $d = 10$, $q = 8$, $B$ vs $4B$, 5 seeds.
- Stage S1 ($d \leq 20$): 4010 runs, 45.85 GPU-h. Stage A10 ($d = 100$; rackwitz_d100 from the frozen plan and series4_d100): 580 runs, 26.33 GPU-h including the $B$ pilot, code checks and the per-mode rebuild. Planned stages S1b ($d = 20, 40$) and S2 (RP63) were not run (Amendment A10); there are no results at $d = 40$. Amendment A10 also reduced four pre-specified parts of the design: (i) Block C was run on four-branch only (pre-specified: four-branch, Rackwitz $d = 20$ and RP54); (ii) Block D was run at $d = 10$ only (pre-specified: $d = 10$ and 40); (iii) at $d = 100$ Block A contained only the six H0-main methods, without POOL-KM-U and CONT-GA-PQPVC, so the H2 comparisons at $d = 100$ choose the best continuous method without GA-PQPVC and the best pool method without POOL-KM-U; (iv) Block B ($q = 1$) at $d = 100$ was run on series4_d100 only, not on Rackwitz $d = 100$.

**Metrics and hypotheses.** The metrics and hypotheses were fixed in the frozen plan.

- Primary metrics: E1 = $|\hat{P}_f - P_f|/P_f$ and E2 = $N_{\text{call}}$, at stop ($d = 100$: after the final refit).
- H0-main (pre-specified; the frozen plan states it as a zero median paired difference): for every (problem, $q \in \{4,8\}$, opponent $\in$ {CONT-LBFGS, CONT-CMAES, CONT-SSO, POOL-KB-U, POOL-IF-U}), SSOLB and the opponent do not differ, for E1 and for E2. It is tested with the two-sided Wilcoxon signed-rank test, whose null hypothesis is that the paired differences SSOLB − opponent are symmetric about zero; the test concerns the distribution of the paired differences, not their median alone.

- H1 (pre-specified, Block A): differences between optimizers do not grow with $d$ (Friedman [68] and Page [69] trend tests).
- H2 (pre-specified, Block A): at $d \geq 20$, the best continuous and the best pool method need the same $N_{\mathrm{call}}$.
- H3 (pre-specified): SSOLB and multistart L-BFGS-B have the same per-round inner time (timing replay on an exclusive GPU, scope reduced by Amendment A12).
- H-MM (pre-specified secondary, Amendments A3, A4 and A6): on series systems, each method vs CONT-LBFGS in (i) the share of runs with a missed mode at stop (McNemar exact [70]), (ii) $N_{\mathrm{call}}$ at full coverage (Wilcoxon), and (iii) the share of runs with an unlearned mode (Amendment A6 (b), McNemar exact).
- Secondary and sensitivity metrics: success rate E4 (the share of runs that reached E1 ≤ 0.05 and did not stop at $N_{\max}$; numbered as in the frozen plan), symmetric log error (Amendment A6 (a)), $\hat{P}_f = 0$ counts, per-mode error share, batch diversity.

### 5.4. Statistics, Families and Reproducibility Controls

**Tests and families.** All pre-specified comparisons are paired over seeds.

- Paired two-sided Wilcoxon signed-rank tests of the 20 paired differences (null hypothesis: symmetric about zero; zero_method "pratt") [71, 72]; Hodges–Lehmann estimate of the paired differences (median of their Walsh averages) with a seed-bootstrap 95% CI ($10^4$ resamples) [73].
- Holm correction [74] within families: H0-main E1 and E2 each 110 tests (S1 90 + Stage A10 20); H-MM 213 tests (S1 171 + Stage A10 42); the Amendment A6 (a) sensitivity family separately.
- W/T/L: W if the Holm-adjusted $p < 0.05$ and the mean paired difference favours SSOLB, L if it favours the opponent, otherwise T (definition of [18]). In a post hoc audit of every significant H0-main cell (46 E1 and 33 E2 cells), the mean paired difference, the Hodges–Lehmann estimate and the direction of the signed-rank statistic (sum of the positive minus sum of the negative ranks) always agreed, so no label depends on which of the three defines the direction.
- TOST [75] with bounds fixed before S1: E1 difference within ±0.02; $N_{\mathrm{call}}$ ratio within [0.90, 1.11] (tested on the log ratio). In S1 the test procedure was chosen after the data were seen, so S1 TOST results are exploratory; for Stage A10 the procedure of Amendment A9 applies. E2 cells in which both methods stopped at $N_{\max}$ in all 20 seeds are censored: neither equivalent nor different (Amendment A9).

- Inner-loop time: runs shared GPUs, so their times are descriptive only; H3 uses a replay of saved GP states on one exclusive GPU (Amendment A12): for each seed 0–4 run of S1 and Stage A10 (1,135 runs, all methods), the GP of the middle round is rebuilt from its saved hyperparameters, and CONT-LBFGS and SSOLB each select a batch from it three times at the run's budget $B$, in random interleaved order; the median of the three times is used. Per (problem, $q$) cell, a two-sided Wilcoxon signed-rank test (zero_method "pratt") on log $\rho$, $\rho = t_{\mathrm{SSOLB}}/t_{\mathrm{LBFGS}}$, Holm over the 32 cells (H3 family); effect sizes are the median $\rho$ and the Hodges–Lehmann estimate of log $\rho$ transformed back, with state-bootstrap 95% CIs ($10^4$ resamples). The analysis was specified before any replay timing existed (Amendment A12).

**Reproducibility controls.** Three controls guard the computations.

- Deterministic GPU settings; 15/15 designed bit-identity pairs within S1 identical; the Stage A10 code version reproduced the S1 version bit for bit in four test runs on rackwitz_d100, series4_d20 and fourbranch_k6 (cluster job 540281, as recorded in the cluster job log; the outputs of this check will be added to the data package before publication).
- Budget audit: in all 233,104 selection rounds of S1, no optimizer exceeded its budget.
- Per-mode estimates $\hat{P}_k$: the final GP of every multi-mode run was rebuilt on the same GPU platform with the original MC population; $\hat{P}_f$ and $\hat{\varepsilon}$ were reproduced bit for bit in 1320/1320 S1 runs and 340/340 Stage A10 runs before $\hat{P}_k$ was used.

**Use of AI tools in the research process.** A generative AI assistant, Claude (Anthropic), used via Claude Code, assisted the author in the research process as well as in the writing (see the Declaration of Generative AI). Under the author's direction it wrote and revised the experiment and analysis code, prepared and submitted the cluster jobs, ran the analyses, produced the figures and tables, drafted the reports and the manuscript text, and searched and summarized the literature. The hypotheses, methods, budgets, metrics, tests, pass/fail thresholds and every amendment were fixed or approved by the author before the affected analyses were run, and every reported result of our runs is produced by code from recorded outputs, except the compute totals and the dates and times, which are taken from the cluster job records and the commit log; values quoted from the literature or from the plan, the amendments and the decision records are cited as such; the code (except the scripts of the post hoc checks added after the external pre-submission reviews: the robustness checks of H2 and H3, the direction audit, the re-evaluation on an independent MC population, the compensation and censoring checks and the stop-reason, intervention and importance-sampling checks), the frozen plan and amendments, the per-run results and the saved GP states are in the data package, and the remaining code, outputs and records cited in the paper will be added to it before publication (Data and Code Availability). The author checked the code, the results and the cited sources and takes responsibility for them.

## 6. Results

### 6.1. Primary Hypothesis H0-main: SSOLB Was Not Better

Table 5 summarizes the 110 pre-specified comparisons per metric (W/T/L by experiment and by opponent, and the cells discussed below), Appendix Table S2 lists all of them with medians, Hodges–Lehmann differences, Holm-adjusted p values and W/T/L labels, and Figure 2 shows the labels as a map.

- E1: W/T/L = 6/64/40 (S1 4/56/30; Stage A10 2/8/10). Split by opponent, SSOLB was 6/41/19 against the three continuous optimizers that maximize the same criterion, Equation (4) (66 optimizer-only cells), and 0/23/21 against the two pool methods, which minimize $U$ over $S$ (44 criterion-changing cells), so its losses do not arise only from a change of criterion (Table 5). The six wins are four against CMA-ES on fourbranch and fourbranch_k6 (S1) and two against plain SSO on rackwitz_d100 (Stage A10). Of the 40 losses, 37 are on single-mode problems (rackwitz_d6 and d10, both 56-bar truss cases, the oscillator at $q = 4$, and rackwitz_d100, where the median E1 of SSOLB was 0.69 and 0.63 at $q = 4$ and 8 against about 0.012 for the pool methods) and three on series4_d100 (a post-freeze addition); 21 are against pool methods and 19 against CONT-LBFGS (7), CMA-ES (6) and plain SSO (6). The five comparisons of one (problem, $q$) reuse the same SSOLB runs, so the W/T/L counts are not independent findings; Holm's procedure controls the family-wise error rate under any dependence, but the counts are best read per (problem, $q$) row of Figure 2.
- E2: W/T/L = 10/77/23 (S1 10/59/21; Stage A10 0/18/2); all ten wins are at $d = 2$ (rackwitz_d2 and fourbranch_k6). 47 of the 110 E2 cells are censored (both methods at $N_{\max}$ in all seeds: S1 31, Stage A10 16). In 53 further cells some runs stopped at $N_{\max}$ (W/T/L 3/27/23 in these cells), and only 10 cells had no run at the cap; the cap enters the test as an observed value although the call count a capped run would have needed is unknown, so paired differences involving capped runs are shrunk towards zero, which tends to make the test conservative and biases the Hodges–Lehmann estimate towards zero.

- Enlarging the family from 90 to 110 tests changed no S1 label.
- Sensitivity to the importance-sampling design of truss56_orig (post hoc; Section 4.1). Without its ten cells, and with the labels of the 110-test family, E1 was 6/62/32 (same criterion 6/39/15; pool methods 0/23/17) and E2 10/67/23; all ten truss56_orig E2 cells were ties, and SSOLB lost eight of its ten truss56_orig E1 cells, four of them against the pool methods.
- Sensitivity to the MC population (post hoc, exploratory). $\hat{P}_f$ and hence E1 are computed on the run's MC population $S$, which the methods of a seed share (Section 4.1), so E1 contains the sampling error of $S$. After the second external pre-submission review, the saved final GPs of 1,500 runs (all methods; seeds 0–4 of S1, of rackwitz_d100 and of Stage A11 (a post-freeze addition), and seeds 0–19 of series4_d100) were evaluated, without refitting, on an independent population of the same size and sampling scheme, again shared by the methods of a seed, giving E1′ (0.41 GPU-hours on one RTX PRO 6000 GPU, against an estimate of at most 0.8 GPU-hours recorded before the run). The new population changed $\hat{P}_f$ by a median of 1.2%. In all 46 significant E1 cells the mean paired difference over the saved seeds had the direction of the published label, both with E1 and with E1′, and 247 of the 275 seed pairs kept their sign; the cells were not retested, since most have five saved seeds. The order of all methods by median E1 within a (stage, problem, $q$) cell was less stable: the median Kendall $\tau$ between the orders by E1 and by E1′ was 0.64 over 35 cells, and the most accurate method was the same in 12 of them. The gap in median E1 between the two most accurate methods of a cell was at most 0.0071 (median 0.0012), below the median per-run change $|E1' - E1|$ of 0.0088 over all runs. Which method of a cell is the most accurate is therefore not resolved by these data, consistent with the many ties.
- Equivalence (TOST): under the procedure fixed before Stage A10 (Amendment A9; bounds from the frozen plan), no E1 cell was equivalent (Holm-adjusted $p_{\mathrm{TOST}} = 1$ in all 20 cells), and two informative E2 cells were equivalent: SSOLB vs POOL-IF-U on series4_d100 at $q = 4$ and $q = 8$ (median $N_{\mathrm{call}}$ ratio 1.00; Holm-adjusted $p_{\mathrm{TOST}}$ 0.033 and $6.2 \times 10^{-4}$ in the 110-test family). In both cells the median $N_{\text{call}}$ of both methods was at the call cap ($N_{\text{call}} = 1101$: the cap $100 + 10d = 1100$ plus the overshoot of the last batch; Hodges–Lehmann differences 1.0 [0, 48] and 0.0 [0, 0]), so the equivalence holds largely at the cap. In S1, where TOST is exploratory, 52 E1 and 49 E2 cells were within the bounds (31 of these E2 cells censored).

- Interpretation. The pre-specified alternative (SSOLB better) is not supported. The hybrid optimizer was mostly tied with gradient-based and population-based continuous optimizers and worse than pool-based batch selection on single-mode problems. A tie (no significant difference) is not equivalence: equivalence was shown only for the two Stage A10 E2 cells above, at the cap; all other ties are inconclusive.

Table 5. Pre-specified H0-main comparisons, SSOLB − opponent: W/T/L by experiment and by opponent, and the cells discussed in the text ($n = 20$ paired seeds per cell; Holm over 110 tests per metric). All 110 cells: Appendix Table S2.

| Comparison | Cells | E1 W/T/L | E1 HL | E2 W/T/L (censored) | E2 HL |
|---|---|---|---|---|---|
| S1, all opponents | 90 | 4/56/30 | 0.0044 | 10/59/21 (31) | — |
| Stage A10, all opponents | 20 | 2/8/10 | 0.036 | 0/18/2 (16) | — |
| All | 110 | 6/64/40 | 0.0077 | 10/77/23 (47) | — |
| Same criterion (CONT-LBFGS, CONT-CMAES, CONT-SSO) | 66 | 6/41/19 | 0.0045 | 6/53/7 (38) | — |
| Pool methods (POOL-KB-U, POOL-IF-U) | 44 | 0/23/21 | 0.022 | 4/24/16 (9) | — |
| CONT-LBFGS | 22 | 0/15/7 | 0.0044 | 2/17/3 (14) | — |
| CONT-CMAES | 22 | 4/12/6 | 0.0054 | 3/18/1 (13) | — |
| CONT-SSO | 22 | 2/14/6 | 0.0034 | 1/18/3 (11) | — |
| POOL-KB-U | 22 | 0/11/11 | 0.022 | 1/11/10 (5) | — |
| POOL-IF-U | 22 | 0/12/10 | 0.022 | 3/13/6 (4) | — |
| four-branch, $q = 4$, CMA-ES | 1 | W | −0.0998 [−0.124, −0.077] | T | 0.0 [−8.0, 0.0] |

| Comparison | Cells | E1 W/T/L | E1 HL | E2 W/T/L (censored) | E2 HL |
|---|---|---|---|---|---|
| four-branch, $q = 8$, CMA-ES | 1 | W | −0.0296 [−0.042, −0.0158] | T | −8.0 [−16.0, −4.0] |
| fourbranch_k6, $q = 4$, CMA-ES | 1 | W | −0.142 [−0.163, −0.121] | T | −4.0 [−8.0, 0.0] |
| fourbranch_k6, $q = 8$, CMA-ES | 1 | W | −0.0377 [−0.0521, −0.0231] | W | −16.0 [−24.0, −12.0] |
| Rackwitz $d = 100$, $q = 4$, SSO | 1 | W | −21.3 [−22.1, −20.4] | Tc | 0.0 [0.0, 0.0] |
| Rackwitz $d = 100$, $q = 8$, SSO | 1 | W | −30.7 [−31.8, −29.5] | Tc | 0.0 [0.0, 0.0] |
| Rackwitz $d = 100$, $q = 4$, KB-U | 1 | L | 0.661 [0.619, 0.684] | Tc | 0.0 [0.0, 0.0] |
| Rackwitz $d = 100$, $q = 8$, KB-U | 1 | L | 0.606 [0.588, 0.625] | Tc | 0.0 [0.0, 0.0] |
| Rackwitz $d = 100$, $q = 4$, IF-U | 1 | L | 0.661 [0.619, 0.684] | Tc | 0.0 [0.0, 0.0] |
| Rackwitz $d = 100$, $q = 8$, IF-U | 1 | L | 0.605 [0.589, 0.625] | Tc | 0.0 [0.0, 0.0] |
| series4_d100, $q = 4$, IF-U | 1 | L | 0.0355 [0.0247, 0.0469] | T | 1.0 [0.0, 48.0] |

| Comparison | Cells | E1 W/T/L | E1 HL | E2 W/T/L (censored) | E2 HL |
|---|---|---|---|---|---|
| series4_d100, $q = 8$, IF-U | 1 | T | 0.0361 [0.0241, 0.0507] | T | 0.0 [0.0, 0.0] |

W/T/L: W if $p_{\mathrm{H}} < 0.05$ and the mean paired difference favours SSOLB (negative), L if it favours the opponent, otherwise T; for single cells (problem, $q$, opponent; CMA-ES: CONT-CMAES, SSO: CONT-SSO, KB-U: POOL-KB-U, IF-U: POOL-IF-U) the label is shown; same criterion: the 66 optimizer-only cells, opponents that maximize the same criterion as SSOLB; pool methods: the 44 criterion-changing cells (CONT-GA-PQPVC, which also changes the criterion, is not an H0-main opponent); c: censored E2 cell (both methods at $N_{\max}$ in all 20 seeds; Amendment A9). HL for groups of cells: the unweighted median over the cells of the per-cell Hodges–Lehmann estimates of the E1 difference (positive: SSOLB less accurate); E2 is not summarized across cells (—), because $N_{\max}$ differs by a factor of about 9 between problems (120 to 1100) and 47 of the 110 cells are censored. HL for single cells: the Hodges–Lehmann estimate with seed-bootstrap 95% CI ($10^4$ resamples); $p_{\mathrm{H}}$, the Holm-adjusted p (110 tests per metric), is listed for every cell, including the single cells shown here, in Appendix Table S2 and is not repeated here. The bootstrap CIs are not adjusted for multiplicity, so a CI that excludes 0 can go with T (for example four-branch, $q = 8$, E2, and series4_d100, $q = 8$, E1); E1 HL is a difference in relative error, E2 HL in limit-state calls. series4_d100 is a post-freeze addition (Section 5.1); its two cells against POOL-IF-U are the two E2 cells equivalent under the TOST procedure fixed before Stage A10 (Amendment A9), both at the call cap.

| | E1 (W/T/L = 6/64/40) | | | | | E2 (W/T/L = 10/77/23) | | | | |
|---|---|---|---|---|---|---|---|---|---|---|
| | LBFGS | CMA-ES | SSO | KB | IF | LBFGS | CMA-ES | SSO | KB | IF |
| four-branch, q = 4 | T | W | T | T | T | L | T | T | T | T |
| four-branch, q = 8 | T | W | T | T | T | T | T | T | T | T |
| fourbranch_k6, q = 4 | T | W | T | T | T | T | T | L | T | T |
| fourbranch_k6, q = 8 | T | W | T | T | T | W | W | T | T | W |
| series4_d20, q = 4 | T | T | T | T | T | Tc | T | Tc | L | T |
| series4_d20, q = 8 | T | T | T | T | T | Tc | Tc | Tc | L | T |
| series4_d100, q = 4 | T | T | L | T | L | Tc | Tc | Tc | L | T |
| series4_d100, q = 8 | T | T | T | L | T | Tc | Tc | Tc | L | T |
| oscillator, q = 4 | L | T | L | L | L | L | Tc | L | L | L |
| oscillator, q = 8 | T | T | T | T | T | L | L | L | L | L |
| Rackwitz d = 2, q = 4 | T | T | T | T | T | T | W | W | W | W |
| Rackwitz d = 2, q = 8 | T | T | T | T | T | W | W | T | T | W |
| Rackwitz d = 6, q = 4 | L | T | L | L | L | Tc | Tc | T | L | L |
| Rackwitz d = 6, q = 8 | L | L | T | L | L | Tc | Tc | Tc | L | L |
| Rackwitz d = 10, q = 4 | L | L | L | L | L | Tc | Tc | Tc | Tc | Tc |
| Rackwitz d = 10, q = 8 | L | L | T | L | L | Tc | Tc | Tc | Tc | Tc |
| Rackwitz d = 100, q = 4 | T | L | W | L | L | Tc | Tc | Tc | Tc | Tc |
| Rackwitz d = 100, q = 8 | L | L | W | L | L | Tc | Tc | Tc | Tc | Tc |
| 56-bar truss, q = 4 | T | T | L | L | L | Tc | Tc | Tc | Tc | T |
| 56-bar truss, q = 8 | L | L | L | L | L | Tc | Tc | Tc | T | T |
| 56-bar truss, mod., q = 4 | T | T | T | L | T | Tc | T | T | L | L |
| 56-bar truss, mod., q = 8 | T | T | T | T | T | Tc | Tc | T | L | L |

W: SSOLB better
L: opponent better
T: no significant difference
Tc: censored E2 cell

Figure 2. W/T/L labels of the 110 pre-specified H0-main comparisons (Appendix Table S2), SSOLB against each opponent, for E1 (left) and E2 (right). W: SSOLB better (Holm-adjusted $p < 0.05$); L: opponent better; T: no significant difference; Tc: censored E2 cell (both methods at $N_{\max}$ in all 20 seeds). Rows: problem and batch size $q$. Columns: opponents (LBFGS: CONT-LBFGS; CMA-ES: CONT-CMAES; SSO: CONT-SSO; KB: POOL-KB-U; IF: POOL-IF-U). The series4_d100 rows are a post-freeze addition (Section 5.1).

**Pre-specified secondary hypotheses H1 and H2, and the Amendment A6 sensitivity analyses.** The operational details of H1 and H2 that the frozen plan left open (the optimizer set, the response of the trend test, the definition of "best") were fixed in the analysis script before any H1 or H2 output was seen (Appendix Table S1); the pre-specified results below use the Block A cells only, with Holm adjustment within each family, and the $q = 1$ cells added at analysis time are exploratory.

- H1 ($N_{\text{call}}$; uninformative). The pre-specified alternative, that the $N_{\text{call}}$ differences between the four continuous inner optimizers grow with $d$, was not supported in any of the four Block A tests (Rackwitz $d \in \{2,6,10,100\}$ and multi-mode $d \in \{2,20,100\}$, with series4_d100 at $d = 100$, each at $q = 4$ and 8; one-sided Page p for an increase ≥ 0.99). The two-sided Page tests indicated a decrease (Holm-adjusted p $4.2 \times 10^{-6}$ to 0.0049), but this is an artefact of censoring: for $d \geq 6$ on the Rackwitz family and $d \geq 20$ on the multi-mode family almost every run stopped at $N_{\text{max}}$, and the Friedman statistic was 0 at $d = 10$ and $d = 100$. In its pre-specified form H1 is therefore uninformative. The two $q = 1$ families added at analysis time showed the same pattern (exploratory). Exploratory, on E1: the spread of E1 between the optimizers grew with $d$ on the Rackwitz family (Holm-adjusted $p \leq 1.7 \times 10^{-13}$ at $q = 4$ and 8), in part because of the plain-SSO failure at $d = 100$ (Section 6.3); on the multi-mode family it did not grow (it decreased at $q = 4$, $p_{\text{H}} = 0.042$, and showed no trend at $q = 8$).

- H2 ($N_{\text{call}}$; rejected, in favour of pool with Kriging-believer selection). Of the six pre-specified Block A cells at $d \geq 20$, four were informative, and in all four the best pool method, POOL-KB-U in each, satisfied the stopping criterion with fewer limit-state calls than the best continuous method (series4_d20 and series4_d100; $q = 4$ and 8): median $N_{\text{call}}$ 193 and 213 against 301 at $d = 20$, and 677 and 713 against 1101 at $d = 100$; Hodges–Lehmann differences +80 to +404 calls; Holm-adjusted $p = 6.6 \times 10^{-4}$ in each cell. In each of these cells 2–4 of the 20 paired differences were zero and all others favoured POOL-KB-U; with zeros and ties the signed-rank test (Pratt zeros) uses its normal approximation, which gave raw p values of $1.1 \times 10^{-4}$ to $1.7 \times 10^{-4}$, and the adjusted values coincide because Holm's step-down procedure does not let an adjusted p value fall below the one before it. POOL-KB-U stopped on $\varepsilon$ in 17–18 of 20 runs, the best continuous methods in none. The result is specific to Kriging-believer selection: in the same four cells the other pool batch methods had their median $N_{\text{call}}$ at the call cap wherever they were run, as every continuous method did (POOL-IF-U 301, 301, 1101 and 1101, stopping on $\varepsilon$ in 1–7 of 20 runs; POOL-KM-U 301 and 301 at $d = 20$, stopping on $\varepsilon$ in none, and not run at $d = 100$, Amendment A10). POOL-KB-U and POOL-IF-U were not tested against each other (both are opponents of SSOLB only), so this ranking of the pool methods is descriptive. The two rackwitz_d100 cells were censored (both methods at $N_{\text{max}}$ in all runs). All informative cells are on the two four-mode series-system problems, and series4_d100 was chosen after S1 (Section 7.3). "Best" was chosen from the same results that the test then compared: the method with the lowest median $N_{\mathrm{call}}$ (ties by median E1, then by name). The difference is in the calls needed to satisfy the surrogate-based stopping criterion $\hat{\varepsilon} \leq 0.02$, not in calls for an equal, verified accuracy ($\hat{\varepsilon}$ can accept underestimates; Section 6.2). E1 did not differ, except in one cell where the pool method was more accurate (rackwitz_d100, $q = 4$; exploratory). Post hoc robustness check (exploratory, specified after the H2 result was known; same tests, Holm over the six cells of each pair): POOL-KB-U satisfied the stopping criterion with fewer limit-state calls than CONT-LBFGS and than SSOLB in the same four informative cells, with the same medians and Hodges–Lehmann differences (+80 to +404 calls; Holm-adjusted $p = 6.6 \times 10^{-4}$ in each cell and for each pair), and both pairs were censored in the two rackwitz_d100 cells. Because the median $N_{\mathrm{call}}$ of every continuous method was at the cap in the four informative cells, the choice of the continuous method does not affect H2; the choice of the pool method does, as stated above. On E1 (exploratory), POOL-KB-U was more accurate than both fixed partners in the four $d = 100$ cells (Holm-adjusted $p \leq 0.036$) and did not differ from them at $d = 20$. The two $q = 1$ cells added at analysis time (AKMCS-U against the best continuous method on series4_d20 and series4_d100) pointed the same way (+106 and +414 calls; exploratory).

- H3 (inner-loop time; rejected, in favour of SSOLB). The 1,135 GP states were replayed on one exclusive H200 GPU; states of one problem and seed share the MC population and the initial design, so the seed is the independent unit. With one value per problem and seed (55 units, the median log $\rho$ over $q$ and source methods; post hoc), the median ratio $\rho = t_{\mathrm{SSOLB}}/t_{\mathrm{LBFGS}}$ was 0.53 and $\rho < 1$ in all 55 units (two-sided sign test $p = 5.6 \times 10^{-17}$), and a bootstrap over the 55 problem–seed clusters gave a 95% CI of 0.46–0.55 for the pooled median $\rho$ of the states (0.49–0.53 when resampled within problems). The pre-specified per-cell tests (two-sided Wilcoxon signed-rank tests over the states of a cell, Holm over the 32 cells) ignore the dependence of states within a seed: the 25–40 states of a cell come from five seeds and share them, so these p values overstate the evidence, and the conclusion rests on the seed-level result above. In these tests SSOLB needed less time per selection round than multistart L-BFGS-B in all 32 (problem, $q$) cells (Holm-adjusted $p = 5.8 \times 10^{-11}$ to $7.3 \times 10^{-4}$), and over all states the median $\rho$ was 0.51 (state-bootstrap 95% CI 0.48–0.54, which also ignores that dependence; Hodges–Lehmann 0.47;

median times 1.10 s against 2.48 s), between 0.23 and 0.78 per cell, and below 1 in 90% of states. The advantage grew with $q$ (median $\rho$ 0.65, 0.49 and 0.40 at $q = 1$, 4 and 8) but not monotonically with $d$. Validity checks: every measurement stayed within $B$; the three repetitions of each optimizer reached identical criterion values; no other process used the GPU; the replayed batch was bit-identical to the stored one in 265 of 320 states from CONT-LBFGS and SSOLB runs (the replay itself was deterministic, and $\rho$ was similar in the 55 other states: 0.51 against 0.56; see Section 7.3 for their cause). Per cell, the seed-level $\rho$ was below 1 for all five seeds in 31 of the 32 cells; with five seeds a sign test reaches at best $p = 0.031$ one-sided (five of five) and $p = 0.0625$ two-sided, so a seed-level version of the two-sided per-cell tests could not reach $p < 0.05$, and a one-sided version would be raised to 1 by the Holm correction over 32 cells; per cell, only the state-level p values above are available. Part of the difference reflects the implementation, since the L-BFGS-B steps run on the CPU and SSO runs entirely on the GPU, and the node's CPUs were shared with other jobs (1-minute load 26–156), which may have slowed L-BFGS-B more than SSOLB; we did not quantify this.

- Amendment A6 (a), symmetric log error L = $|\log10(\hat{P}_f / P_f)|$, Stage A10. In the 110-test sensitivity family the conclusions agreed with E1 in 19 of the 20 Stage A10 cells (13 significant against 12 for E1, all in the same direction), and the 34 significant S1 tests were unchanged by the enlargement. The exception is rackwitz_d100 at $q = 4$ against CONT-LBFGS: a tie on E1 ($p_{\mathrm{H}} = 0.17$) but a loss for SSOLB on L (median 0.51 against 0.43; $p_{\mathrm{H}} = 0.026$). Under Amendment A6 the E1 result is the pre-specified one. $\hat{P}_f = 0$ was substituted by $0.5/N_{\mathrm{MC}}$ in four CMA-ES runs on series4_d100.
- Amendment A6 (b2), Stage A10. With the interval [1/3, 3] instead of [0.5, 2], all 14 decisions on series4_d100 were the same. At $q = 1$ CONT-LBFGS left a mode unlearned in 12/20 runs (14/20 with (b)) against 0/20 for each of the other four methods (McNemar $p = 4.9 \times 10^{-4}$; outside the Holm families).

### 6.2. Secondary Hypothesis H-MM: Missed and Unlearned Modes and Their Safety Consequence

Family: 213 tests (S1 171 + Stage A10 42), Holm.

The learned-mode criterion (Amendment A6 (b)), on which the unlearned-mode tests and counts below rest, was fixed after Stage S0, whose pre-tests ran CONT-LBFGS only, and before S1 (Appendix Table S1). In two S0 pilot runs on four-branch at $q = 1$, modes counted as covered by the distance criterion of Amendment A3 had $\hat{P}_k = 0$, which showed that coverage does not establish that the probability of a mode is estimated; no comparison between methods had been run. The criterion is therefore pre-specified with respect to S1 and Stage A10, but it was chosen after outputs of the method it most affects had been seen; with [1/3, 3] instead of [0.5, 2] (Amendment A6 (b2)) the decisions were almost unchanged (Sections 6.1 and 7.2).

- $q = 1$, S1 ($d = 2$ and 20). CONT-LBFGS stopped with a never-sampled mode in 19/20 (fourbranch), 12/20 (fourbranch_k6) and 14/20 (series4_d20) runs, and with an unlearned mode (Amendment A6 (b)) in 20/20, 20/20 and 19/20 runs (Figure 3). In the 213-test family, 38 tests were significant (S1 34, Stage A10 4; Appendix Table S3). At $q = 1$ in S1 there were 27, and all of them favoured the other method over CONT-LBFGS: fewer runs with a missed mode for GA, SSO and AKMCS-U on four-branch and for CMA-ES, GA, SSO, SSOLB and AKMCS-U on series4_d20 (8 tests); fewer runs with an unlearned mode for GA, SSO and AKMCS-U on four-branch, for GA, SSO, SSOLB and AKMCS-U on fourbranch_k6 and for CMA-ES, GA, SSO, SSOLB and AKMCS-U on series4_d20 (12); and fewer calls to full coverage for GA and SSO on four-branch and for CMA-ES, GA, SSO, SSOLB and AKMCS-U on series4_d20 (7; Hodges–Lehmann −58 to −240 calls).
- $q = 1$, Stage A10 (series4_d100, a post-freeze addition). Coverage alone did not separate methods: 7/20 initial designs ($n_0 = 101$) already covered all four modes, and CONT-LBFGS missed a mode in 1/20 runs; no coverage test was significant. The learned-mode criterion did separate them: CONT-LBFGS left a mode unlearned in 14/20 runs (mostly the smallest mode, $P_4 = 2.3 \times 10^{-4}$, learned in 6/20 runs, median $\hat{P}_4/P_4 = 0.22$), against 0/20 for SSOLB, SSO, GA and AKMCS-U; McNemar $p = 1.2 \times 10^{-4}$, Holm-adjusted $p = 0.023$ for each of the four comparisons. These values come from the bit-identical GPU rebuild of the final GP states (Section 5.4); a CPU approximation on an independent MC population gave the same classification for all 340 runs.
- $q = 4$ and 8. Seven S1 tests were significant in the 213-test family, six of them in favour of CONT-LBFGS. GA-PQPVC left a mode unlearned on four-branch at $q = 8$ in 17/20 runs against 0/20 for CONT-LBFGS ($p_{\mathrm{H}} = 0.0024$ in the S1-only family and 0.0031 in the 213-test family), and it needed more calls to full coverage on four-branch and fourbranch_k6 at $q = 4$ and 8 (Hodges–Lehmann +16 to +78 calls); POOL-KB-U needed more calls to full coverage on four-branch at $q = 4$ (+36). Plain SSO needed fewer on series4_d20 at $q = 4$ (−26). No missed-mode test was significant at $q \geq 4$. In Stage A10, CMA-ES at $q = 8$ had 5/20 runs with an unlearned mode (vs 0/20, $p = 0.062$, not significant after Holm).

- Family enlargement. One S1 test lost significance when the family grew from 171 to 213: the full-coverage $N_{\mathrm{call}}$ test for SSOLB vs CONT-LBFGS on fourbranch at $q = 8$ ($p_{\mathrm{H}}$ 0.045 → 0.057). Significant S1 H-MM tests: 35 → 34.
- Direction. The pre-specified hypothesis predicted that population-based optimizers miss fewer modes than multistart L-BFGS-B. At $q = 1$ this held for plain SSO and GA-PQPVC on all three S1 problems, for SSOLB on fourbranch_k6 and series4_d20, and for CMA-ES only on series4_d20: CMA-ES left a mode unlearned in 20/20 four-branch and 18/20 fourbranch_k6 runs (39 of its 60 S1 runs at $q = 1$). SSOLB left a mode unlearned in 15/20 four-branch runs (CONT-LBFGS 20/20; McNemar $p = 0.062$, not significant), against 1/20 on fourbranch_k6 and 0/20 on series4_d20. It did not hold at $q \geq 4$, where the batch itself spreads the points and most significant differences favoured CONT-LBFGS over GA-PQPVC.

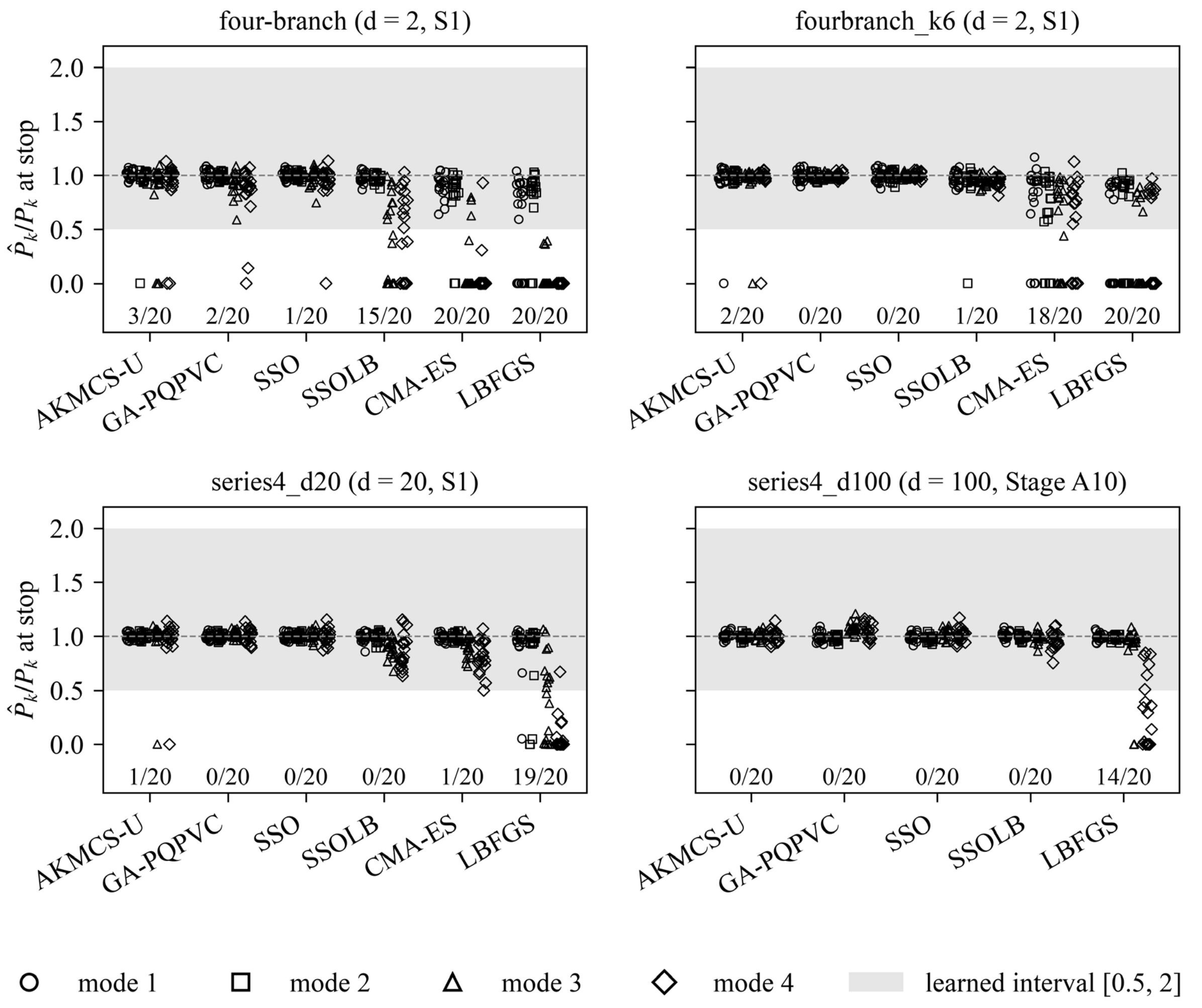

Figure 3. Per-mode accuracy at stop with sequential selection ($q = 1$): the ratio $\hat{P}_k/P_k$ of every mode $k$ in every run, by method, on the three S1 multi-mode problems and on series4_d100 (a post-freeze addition; 20 seeds per method; four modes per run). The band [0.5, 2] is the "learned" interval of Amendment A6 (b); points below it are unlearned modes. Markers distinguish the modes; points are jittered horizontally. Numbers below the points: runs with at least one unlearned mode. CMA-ES was not run at $q = 1$ on series4_d100.

**Safety consequence: unlearned modes were associated with $\hat{P}_f$ below $P_f$ (exploratory).** Table 6 groups the Block A and Block B runs on the three S1 multi-mode problems and the 340 Stage A10 runs on series4_d100 (the mode-classified runs; Block C was not classified) by whether at least one mode was unlearned at stop.

- In all 198 runs (Table 6) with an unlearned mode, $P_f$ was underestimated: S1 $q = 1$ 123/123, S1 $q = 4$ and 8 53/53, Stage A10 22/22 (all on series4_d100; Figure 4). This is an observed association, conditional on the estimated mode probabilities. Of these 198 runs, 17 met the error criterion $\hat{\varepsilon} \leq 0.02$ with $\hat{P}_f/P_f$ between 0.38 and 0.89 (S1 $q = 1$: 5; S1 $q = 4$ and 8: 12; Stage A10: 0; seven distinct problem–seed pairs; 13 of the 17 were pool-method runs, POOL-KB-U 6, AKMCS-U 4 and POOL-IF-U 3), so the error criterion can accept substantial underestimation: in these 17 runs it accepted estimates 11–62% below $P_f$; the other 181 exhausted their budget ($N_{\max}$) without meeting the criterion (Table 6). Of the 1,462 runs with all modes learned, 470 stopped by $\hat{\varepsilon}$ and 992 at $N_{\max}$. Because $\sum_k \hat{P}_k = \hat{P}_f$ (Section 2), a mode estimated too low lowers $\hat{P}_f$ unless errors in the other modes compensate, and they could (through overestimated modes or false-failure mass assigned to them). Every unlearned mode in the 198 runs lay below 0.5 (largest $\hat{P}_k/P_k$ 0.499) and no mode in any run exceeded 2; this is an observation, not implied by Amendment A6 (b), and the grouping therefore selects runs with at least one low $\hat{P}_k$. The informative part of the finding is that the learned modes did not compensate (Table 7; post hoc). In the 190 runs with at least one learned mode, the learned modes had a median $\hat{P}_k/P_k$ of 0.956 (10th to 90th percentile 0.776–1.028), and the compensation ratio $C = \sum_{\text{learned}} \hat{P}_k \,/\, \sum_{\text{learned}} P_k$ had a median of 0.950, against 0.989–1.011 (C medians) in the runs with all modes learned; C exceeded 1 in 35 runs, by at most 6.5%, and 8 runs had no learned mode. Among runs with all modes learned, 62–66% had $\hat{P}_f/P_f < 1$ in S1, and the medians were 0.989–0.992 (S1) and 0.995–1.011 (Stage A10).

- Magnitudes, $q = 1$:
  - S1 pooled: median $\hat{P}_f/P_f$ 0.674 [IQR 0.415, 0.808] and median $\Delta\beta$ +0.123 with an unlearned mode, vs 0.989 and +0.004 without. Worst case: series4_d20, CONT-LBFGS, $\hat{P}_f/P_f = 0.059$, $\Delta\beta$ = +0.835.
  - Stage A10 (series4_d100; $d = 100$): median 0.927 [0.917, 0.941], $\Delta\beta$ +0.025 (worst 0.73, $\Delta\beta$ +0.10) with an unlearned mode, vs 0.995 and +0.002 without. All 20 CONT-LBFGS runs at $q = 1$ underestimated $P_f$ (median 0.938).
- $q = 4$ and 8 at $d = 100$: four CMA-ES runs ended with $\hat{P}_f = 0$ (series4_d100; $q = 4$ seed 10; $q = 8$ seeds 7, 10, 18), i.e. "no failure" for a problem with $P_f = 3.5 \times 10^{-3}$; under Amendment A1 these runs could only stop at $N_{\max}$.
- The bias had one sign in every observed case: the unlearned mode had lost failure mass, and in no run had it gained enough to be unlearned from above. At $q = 1$ the effect was smaller at $d = 100$ (series4_d100) than at $d \leq 20$ (median $\Delta\beta$ +0.025 against +0.123), consistent with the unlearned mode at $d = 100$ being mostly the smallest one ($P_4/P_f = 6.6\%$); at $q = 4$ and 8 at $d = 100$ the eight unlearned runs had a median $\widehat{P}_f/P_f$ of 0.000.

**Scope of the safety finding.** The finding has two layers. (i) On benchmarks whose failure modes are known and separable, as here, mode coverage and per-mode accuracy can be measured, and the under-estimation that follows an unlearned mode can be observed directly (Table 6). (ii) For systems whose failure modes are not known in advance, these measurements are not available; the result then motivates coverage diagnostics of the selected points or independent validation samples, which this paper does not validate, and it does not show how often modes go unlearned in such systems.

Table 6. Safety consequence of unlearned modes (exploratory): stop reason, $\hat{P}_f/P_f$ and $\Delta\beta = \hat{\beta} - \beta_{\text{ref}}$ at stop, runs grouped by whether at least one mode was unlearned (Amendment A6 (b): $\hat{P}_k/P_k$ outside [0.5, 2]).

| Stage, $q$ | Problem | Group | Runs | Stopped by $\hat{\varepsilon}$ / at $N_{\max}$ | $\hat{P}_f/P_f$ median [IQR] | Worst $\hat{P}_f/P_f$ | Share $\hat{P}_f/P_f < 1$ | $\Delta\beta$ median | Worst $\Delta\beta$ | $\hat{P}_f = 0$ |
|---|---|---|---|---|---|---|---|---|---|---|
| S1, $q = 1$ | four-branch | ≥ 1 unlearned | 61 | 3 / 58 | 0.732 [0.572, 0.838] | 0.289 | 100% | +0.098 | +0.375 | 0 |
| S1, $q = 1$ | four-branch | none unlearned | 59 | 34 / 25 | 0.990 [0.971, 1.004] | 0.904 | 66% | +0.003 | +0.032 | 0 |
| S1, $q = 1$ | fourbranch_k6 | ≥ 1 unlearned | 41 | 1 / 40 | 0.455 [0.375, 0.605] | 0.183 | 100% | +0.258 | +0.535 | 0 |
| S1, $q = 1$ | fourbranch_k6 | none unlearned | 79 | 50 / 29 | 0.990 [0.972, 1.004] | 0.820 | 68% | +0.004 | +0.067 | 0 |
| S1, $q = 1$ | series4_d20 | ≥ 1 unlearned | 21 | 1 / 20 | 0.764 [0.746, 0.865] | 0.059 | 100% | +0.089 | +0.835 | 0 |

| Stage, $q$ | Problem | Group | Runs | Stopped by $\hat{\varepsilon}$ / at $N_{\max}$ | $\hat{P}_f/P_f$ median [IQR] | Worst $\hat{P}_f/P_f$ | Share $\hat{P}_f/P_f < 1$ | $\Delta\beta$ median | Worst $\Delta\beta$ | $\hat{P}_f = 0$ |
|---|---|---|---|---|---|---|---|---|---|---|
| S1, $q = 1$ | series4_d20 | none unlearned | 99 | 18 / 81 | 0.989 [0.980, 1.004] | 0.886 | 64% | +0.004 | +0.040 | 0 |
| S1, $q = 1$ | all three | ≥ 1 unlearned | 123 | 5 / 118 | 0.674 [0.415, 0.808] | 0.059 | 100% | +0.123 | +0.835 | 0 |
| S1, $q = 1$ | all three | none unlearned | 237 | 102 / 135 | 0.989 [0.973, 1.004] | 0.820 | 66% | +0.004 | +0.067 | 0 |
| S1, $q = 4$ and 8 | all three | ≥ 1 unlearned | 53 | 12 / 41 | 0.801 [0.770, 0.844] | 0.172 | 100% | +0.072 | +0.540 | 0 |
| S1, $q = 4$ and 8 | all three | none unlearned | 907 | 304 / 603 | 0.992 [0.976, 1.006] | 0.747 | 62% | +0.003 | +0.096 | 0 |
| Stage A10, $q = 1$ | series4_d100 | ≥ 1 unlearned | 14 | 0 / 14 | 0.927 [0.917, 0.941] | 0.732 | 100% | +0.025 | +0.102 | 0 |

| Stage, $q$ | Problem | Group | Runs | Stopped by $\hat{\varepsilon}$ / at $N_{\max}$ | $\hat{P}_f/P_f$ median [IQR] | Worst $\hat{P}_f/P_f$ | Share $\hat{P}_f/P_f < 1$ | $\Delta\beta$ median | Worst $\Delta\beta$ | $\hat{P}_f = 0$ |
|---|---|---|---|---|---|---|---|---|---|---|
| Stage A10, $q = 1$ | series4_d100 | none unlearned | 86 | 19 / 67 | 0.995 [0.987, 1.007] | 0.949 | 65% | +0.002 | +0.018 | 0 |
| Stage A10, $q = 4$ and 8 | series4_d100 | ≥ 1 unlearned | 8 | 0 / 8 | 0.000 [0.000, 0.179] | 0.000 | 100% | +∞ | ∞ | 4 |
| Stage A10, $q = 4$ and 8 | series4_d100 | none unlearned | 232 | 45 / 187 | 1.011 [0.994, 1.036] | 0.946 | 36% | −0.004 | +0.019 | 0 |

Stopped by $\hat{\varepsilon}$: $\hat{\varepsilon} \leq 0.02$ in two consecutive rounds (Amendment A1); at $N_{\max}$: the call cap (post hoc count). $\hat{P}_f = 0$ gives $\hat{\beta} = \infty$; the median $\Delta\beta$ of +∞ in the Stage A10 $q = 4$ and 8 group arises because 4 of its 8 runs ended with $\hat{P}_f = 0$. Per-mode estimates from the bit-identical rebuild of the final GP states (Section 5.4). The Stage A10 rows are series4_d100, a post-freeze addition (Section 5.1).

Table 7. Compensation by the learned modes in the 198 runs with an unlearned mode (post hoc, exploratory): the compensation ratio $C = \sum_{\text{learned}} \hat{P}_k \,/\, \sum_{\text{learned}} P_k$ by the groups of Table 6.

| Stage, $q$ | Runs with an unlearned mode (with a learned mode) | C median | Runs with C > 1 |
|---|---|---|---|
| S1, $q = 1$ | 123 (122) | 0.928 | 10 |
| S1, $q = 4$ and 8 | 53 (52) | 0.987 | 20 |
| Stage A10, $q = 1$ | 14 (14) | 0.993 | 4 |
| Stage A10, $q = 4$ and 8 | 8 (2) | 0.923 | 1 |
| All | 198 (190) | 0.950 (0.642–1.065) | 35 |

A learned mode has $\hat{P}_k/P_k$ in [0.5, 2] (Amendment A6 (b)); C is computed over the learned modes of a run, with the mode-wise references $P_k$, whose sum differs from the reference $P_f$ by up to 0.1%. All: C median with its range over the 190 runs with a learned mode (the pooled $\hat{P}_k/P_k$ of the learned modes and the largest $\hat{P}_k/P_k$ of an unlearned mode are in Section 6.2). C is undefined in the 8 runs without a learned mode, 6 of them in the Stage A10 $q = 4$ and 8 group (4 of its runs ended with $\hat{P}_f = 0$). Per-mode estimates from the bit-identical rebuild of the final GP states, as in Table 6 (Section 5.4). The Stage A10 rows are series4_d100, a post-freeze addition (Section 5.1).

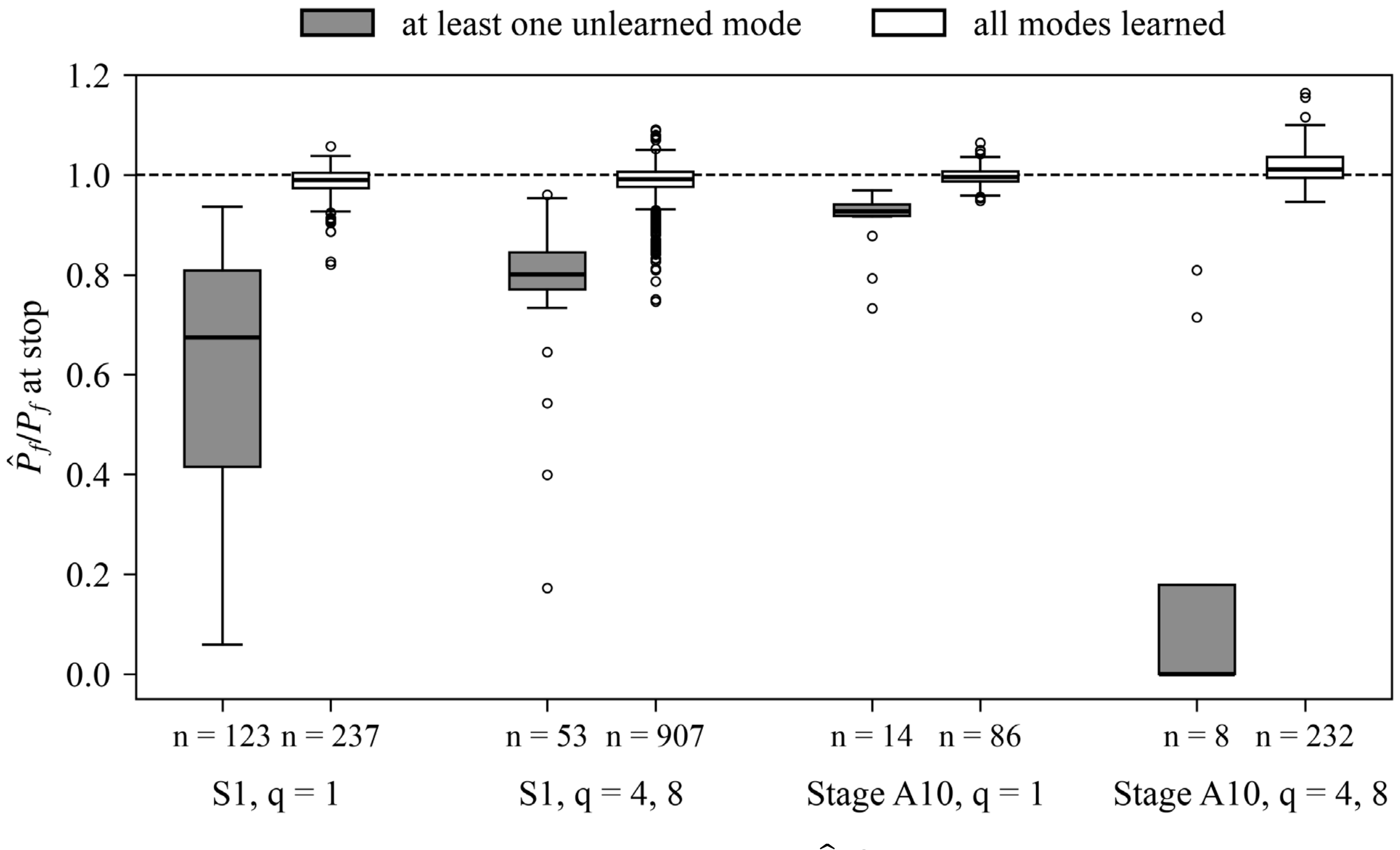

Figure 4. Estimated over reference failure probability $\hat{P}_f/P_f$ at stop for runs with at least one unlearned mode and runs with all modes learned (Amendment A6 (b)), for the Block A and Block B runs on the three S1 multi-mode problems and the 340 Stage A10 runs on series4_d100, a post-freeze addition (the mode-classified runs; Block C was not classified), by stage and batch size (exploratory; the groups of Table 6). Boxes: quartiles and median; whiskers: 1.5 interquartile ranges; dots: runs beyond them. The dashed line marks an unbiased estimate; values below it are unconservative. n: number of runs per group.

### 6.3. Exploratory Comparisons: Candidate-Pool Selection, Plain SSO and Controls

**Candidate-pool selection at $q = 1$.** The pool method AKMCS-U was compared with the continuous methods in exploratory families.

- S1 (nine problems, $d = 2$–$20$; Holm within 117-test exploratory families). AKMCS-U was never worse than plain SSO or SSOLB on E1 or E2: SSO vs AKMCS-U W/T/L 0/8/1 (E1) and 0/5/4 (E2); SSOLB vs AKMCS-U 0/5/4 and 0/2/7.
- S1 had no comparison of AKMCS-U with CONT-LBFGS, CMA-ES or GA-PQPVC.
- Stage A10 (series4_d100, a post-freeze addition; Holm over four comparisons). AKMCS-U (median E1 0.010) did not differ significantly from SSO, SSOLB and GA (medians 0.010, 0.010 and 0.015; all T) and was more accurate than CONT-LBFGS (0.062; W). It needed fewer calls to the stopping criterion than all four continuous methods (median 688 vs 1100; Holm-adjusted $p = 3.8 \times 10^{-4}$ for each), stopping on $\varepsilon$ in 19/20 runs. Its inner time per round was 0.94× that of SSO and 0.11× that of SSOLB (descriptive; shared GPUs).
- Pool methods are not immune to missing modes: in S1 at $q = 1$, AKMCS-U stopped with a never-sampled mode in 3/20 (fourbranch), 1/20 (fourbranch_k6) and 1/20 (series4_d20) runs; in Stage A10 (series4_d100), POOL-KB-U ($q = 4$) and POOL-IF-U ($q = 8$) each had one run with an unlearned mode.

**Plain SSO at $d = 100$.** Holm within 24 tests per metric; SSO − opponent.

- series4_d100: plain SSO reached a median E1 of 0.010, 0.011 and 0.021 at $q = 1, 4$ and 8, and won on E1 against CONT-LBFGS ($q = 1, 4$) and SSOLB ($q = 4, 8$; Holm-adjusted p 0.001 and 0.026). The SSOLB comparisons use the same paired runs as the H0-main cells SSOLB vs plain SSO, where they are an SSOLB loss at $q = 4$ ($p_{\mathrm{H}} = 0.0062$) and a tie at $q = 8$ ($p_{\mathrm{H}} = 0.144$; Figure 2): the difference is the size of the family (24 against 110 tests).
- rackwitz_d100: plain SSO failed. Median E1 21.6 ($q = 4$) and 31.1 ($q = 8$), i.e. $\hat{P}_f$ 23–32 times the reference. Its selected points lay far outside the failure region, at ‖ $u \| \approx 52$–$60$ against $\beta \approx 2.9$ (box half-width $R = 14.8$; uniform points in the box have $\| u \| \approx 85$). SSO lost on E1 to all five opponents at $q = 4$ and 8.

- Loss cells (all named): E1 — rackwitz_d100 $q = 4$ and $q = 8$ vs CONT-LBFGS, SSOLB, CMA-ES, POOL-KB-U and POOL-IF-U (10 cells); E2 — series4_d100 $q = 4$ and $q = 8$ vs POOL-KB-U, $q = 1$ vs AKMCS-U (3 cells).
- These results motivate a separate study of improved swarm search in the inner loop; they are not a claim of this paper.

**Negative control and budget sensitivity.** The two exploratory blocks gave the following results.

- Block C (four-branch, $q = 4$, 5 seeds per arm; all six arms): continuous search on raw $U$ stopped at $N_{\max}$ in 5/5 runs for both optimizers (median E1 0.032 with CONT-LBFGS and 0.009 with SSOLB), while continuous search on the density-weighted criterion stopped by $\varepsilon$ in 4/5 runs for each optimizer (median E1 0.007 and 0.009), consistent with the degeneracy argument of Section 4.1. With pool selection the pattern was reversed: on raw $U$, 4/5 runs stopped by $\varepsilon$ (median E1 0.009), whereas on the density-weighted criterion all 5 stopped at $N_{\max}$ and $P_f$ was underestimated (median E1 0.110, median $\hat{P}_f / P_f$ 0.89), which may reflect the concentration mechanism of Section 6.4 (shown there only on single-design-point problems) or under-learning of modes (four-branch has four); the five seeds do not separate the two (post hoc).
- Block D (rackwitz_d10, $q = 8$, 5 seeds): with $4B$ instead of $B$ all runs still stopped at $N_{\max}$; median E1 changed in both directions (CONT-LBFGS used a median 22% of $4B$). No conclusion.

### 6.4. Engineering Demonstration and the Concentration Mechanism (Stage A11, Post-Freeze Addition, Exploratory)

All results of this section are exploratory and come from post-freeze additions (Section 5.1): the case was added after S1 and Stage A10 and lies outside every pre-specified family (Holm adjustment within the case only). The limit state appears to have a single design point (Section 5.2), so the case shows accuracy and call count only and cannot test H-MM; the safety evidence remains with the multi-mode benchmarks (Section 6.2).

- Completion: 440/440 runs (22 method–q arms of nine methods, 20 seeds each). All runs stopped at the call cap $N_{\max} = 200$ ($N_{\text{call}} = 204$ at $q = 8$) and none by $\varepsilon$; E2 is therefore identical across methods and uninformative (all 31 E2 comparisons censored), and the E4 success rate is 0 for every arm. No run returned $\hat{P}_f = 0$. No initial design contained a failure point.

- Accuracy (Table 8): at $q = 4$ and 8, POOL-KB-U, POOL-IF-U and GA-PQPVC had median E1 of 0.010–0.014, and CMA-ES, CONT-LBFGS and plain SSO ($q = 4$) 0.016–0.042; POOL-KM-U was least accurate among the pool methods (0.131 and 0.082). At $q = 1$, AKMCS-U, GA-PQPVC and plain SSO had median E1 of 0.011, 0.013 and 0.018, with $\hat{P}_f/P_{\text{ref}}$ close to 1. CONT-LBFGS, SSOLB and CMA-ES underestimated $P_f$ in 18/20, 20/20 and 20/20 runs (median $\hat{P}_f/P_{\text{ref}}$ 0.92, 0.73 and 0.31; worst $\Delta\beta$ +0.07, +0.17 and +0.42).
- SSOLB (negative result): SSOLB was less accurate than four of the five H0-main opponents at $q = 4$ and at $q = 8$ (E1 losses after Holm within the case; tied with plain SSO at both $q$). Its median E1 was 0.067 ($q = 4$) and 0.075 ($q = 8$), against 0.011–0.026 and 0.011–0.042 for the four opponents it lost to.
- Plain SSO at $q = 8$ overestimated $P_f$ in 20/20 runs (median $\hat{P}_f/P_{\text{ref}}$ 1.08, range 1.04–1.15). The error is in the conservative direction ($\Delta\beta$ between −0.011 and −0.043), but it is a loss in E1 against CMA-ES, GA-PQPVC, POOL-KB-U and POOL-IF-U.
- Mechanism of the $q = 1$ underestimation. A diagnosis whose verdict rules were fixed in advance found no implementation fault. The budgets were used, and CMA-ES, SSOLB and CONT-LBFGS reached the best-found reference maximizer of the density-weighted criterion of Equation (4) in 96%, 88% and 84% of 25 replayed states each (seeds 0–4, five rounds from the first to the last). For each state the reference maximizer is the best point found by screening $2 \times 10^6$ independent N(0, I) points and running L-BFGS-B (at most 1000 iterations, no budget limit) from the 20 best of them and from the point the optimizer returned; an optimizer counted as having reached it when its criterion value was within 0.1 (log units) of the reference value. The reference is the best point found, not a proven global maximum. By the rules fixed in advance the mechanism was "unresolved". Post hoc, we observed that the best-found reference maximizer of Equation (4) lies next to the FORM design point: CMA-ES, SSOLB and CONT-LBFGS placed their points at a median distance of 0.37–0.78 from $u^*$, against 3.48 for AKMCS-U. The missing probability was almost entirely false-safe mass far from $u^*$ (post hoc).

- Two follow-up analyses of existing runs were specified after these observations and before their outputs were generated, to test that mechanism; both passed their thresholds. (i) An intervention on reconstructed truss states, taken from the same Stage A11 runs in which the concentration had been observed, added one point to each state. The states are rounds 0, 47, 94, 140 and 187 of the $q = 1$ runs of CONT-CMAES, SSOLB, CONT-LBFGS and AKMCS-U with seeds 0–4 (20 runs, 100 states); the four round-0 states of a seed are identical (shared initial design), so 85 states are distinct. With the seed as the unit (post hoc), adding the AKMCS-U pool choice reduced the false-safe mass more than adding the best-found reference maximizer of Equation (4) (found with the polishing settings of the diagnosis, from 1,647,185 independent points and L-BFGS-B from the 20 best) in all five seeds (median difference per seed; one-sided sign test $p = 0.031$, the smallest one-sided value five seeds allow; two-sided it would be 0.0625). At the state level the pool choice was better in 56 of the 73 distinct non-tied states (0.77; one-sided sign test $p = 2.6 \times 10^{-6}$, which ignores the dependence of states within a seed, since they share its MC population and initial design, and is therefore descriptive); the pre-specified state-level test, which also ignores that dependence and counts the identical round-0 states of a seed four times, gave 62 of 79 non-tied states (0.78; threshold 0.60; one-sided sign test $p = 1.8 \times 10^{-7}$). The conclusion rests on the seed-level result. (ii) A cross-problem check using existing runs used the saved selections of all Stage S1 and Stage A10 single-design-point problems (rackwitz $d = 2$, 6, 10, 100; oscillator; truss56_mod; truss56_orig; 20 problem–q cells). Methods whose points lay farther from the design point had higher $\hat{P}_f / P_{\text{ref}}$ in 16 of the 20 cells: median Spearman $\rho = 0.86$ (threshold 0.30), and $\rho > 0$ in 16/20 cells (threshold 0.60). The four exceptions were the two 56-bar truss problems at $q = 4$ and 8, where no method's median $\hat{P}_f / P_{\text{ref}}$ was below 0.98. Both analyses therefore support the mechanism, with three qualifications: they were specified after Stage A11, the intervention used the same truss runs in which the concentration had been observed, and the cross-problem check is correlational. On the 23-bar truss at $q = 1$, the optimizers that reached the best-found reference maximizer of Equation (4) placed their points closest to $u^*$ and underestimated $P_f$ most, through false-safe mass elsewhere in the failure domain; this ordering is an observed tendency, not a rule, and no analysis measured how well each optimizer maximized Equation (4) on the other problems. On rackwitz_d100, for example, CMA-ES, whose points lay closest to $u^*$ on the truss, placed its points far from $u^*$ (median distance 8.2 and 12.0 at $q = 4$ and 8) and had $\hat{P}_f/P_{\text{ref}}$ of 0.93 and 1.01. In the same cells SSOLB and CONT-LBFGS placed their points closest to $u^*$ (median distance 0.62–1.27, against about 10.4 for POOL-KB-U and POOL-IF-U) and underestimated $P_f$ most (median $\hat{P}_f/P_{\text{ref}}$ 0.31 and 0.37 at $q = 4$, 0.37 and 0.51 at $q = 8$, against 0.99 for the two pool methods); these are the largest underestimates among the pre-specified single-design-point cells ($q = 4$ and 8). The observation that motivated these tests remains post hoc. The mechanism concerns selection concentrated near $u^*$, however it arises, not continuous search as such: post hoc, the pool batch method POOL-KM-U also placed its points close to $u^*$ on rackwitz_d10 (median distance 1.54 at $q = 4$ and 1.97 at $q = 8$, against 3.51–3.68 for POOL-KB-U and POOL-IF-U) and also underestimated $P_f$ there (median $\hat{P}_f/P_{\text{ref}}$ 0.77 and 0.85). In a post hoc sensitivity analysis requested after the result of the cross-problem check was seen, the association did not depend on plain SSO, whose large overestimates at $d = 100$ contribute to the rackwitz_d100 cells, or on rackwitz_d100 itself. Without plain SSO, the median $\rho$ was 0.79 with $\rho > 0$ in 15/20 cells; the additional negative cell was rackwitz_d2 at $q = 8$ ($\rho = -0.25$, previously 0.12). Without rackwitz_d100, the median $\rho$ was 0.85 with $\rho > 0$ in 14/18 cells. These figures are reported alongside, not in place of, the pre-specified result.

Table 8. Stage A11, 23-bar truss (post-freeze addition; exploratory; 20 seeds per arm; $P_{\mathrm{ref}} = 1.5177 \times 10^{-3}$, $\beta_{\mathrm{ref}} = 2.964$). All runs stopped at $N_{\max}$ ($N_{\mathrm{call}} = 200$; 204 at $q = 8$).

| $q$ | Method | E1 median [IQR] | $\hat{P}_f/P_{\mathrm{ref}}$ median | $\hat{P}_f/P_{\mathrm{ref}}$ range | Share $\hat{P}_f/P_{\mathrm{ref}} < 1$ | $\Delta\beta$ median | Worst $\Delta\beta$ |
|---|---|---|---|---|---|---|---|
| 1 | AKMCS-U | 0.011 [0.009, 0.020] | 1.008 | 0.952–1.042 | 35% | −0.003 | +0.015 |
| 1 | CONT-GA-PQPVC | 0.013 [0.009, 0.025] | 1.007 | 0.954–1.053 | 40% | −0.002 | +0.015 |
| 1 | CONT-LBFGS | 0.076 [0.021, 0.096] | 0.924 | 0.799–1.022 | 90% | +0.024 | +0.068 |
| 1 | CONT-CMAES | 0.693 [0.621, 0.747] | 0.307 | 0.235–0.478 | 100% | +0.347 | +0.421 |
| 1 | CONT-SSO | 0.018 [0.008, 0.022] | 1.001 | 0.950–1.030 | 50% | 0.000 | +0.016 |
| 1 | SSOLB | 0.272 [0.208, 0.344] | 0.728 | 0.575–0.904 | 100% | +0.096 | +0.166 |
| 4 | POOL-KB-U | 0.011 [0.005, 0.018] | 1.004 | 0.959–1.040 | 35% | −0.001 | +0.013 |
| 4 | POOL-IF-U | 0.012 [0.008, 0.019] | 1.006 | 0.961–1.037 | 35% | −0.002 | +0.012 |
| 4 | POOL-KM-U | 0.131 [0.106, 0.177] | 0.869 | 0.755–1.025 | 90% | +0.043 | +0.085 |
| 4 | CONT-GA-PQPVC | 0.014 [0.009, 0.024] | 0.989 | 0.962–1.040 | 70% | +0.004 | +0.012 |
| 4 | CONT-LBFGS | 0.026 [0.015, 0.047] | 0.974 | 0.926–1.025 | 90% | +0.008 | +0.024 |
| 4 | CONT-CMAES | 0.024 [0.014, 0.055] | 0.976 | 0.912–1.000 | 100% | +0.007 | +0.028 |
| 4 | CONT-SSO | 0.024 [0.010, 0.033] | 1.019 | 0.967–1.082 | 30% | −0.006 | +0.010 |
| 4 | SSOLB | 0.067 [0.025, 0.135] | 0.933 | 0.786–1.014 | 85% | +0.021 | +0.073 |
| 8 | POOL-KB-U | 0.011 [0.009, 0.020] | 1.007 | 0.955–1.040 | 40% | −0.002 | +0.014 |
| 8 | POOL-IF-U | 0.011 [0.007, 0.018] | 1.006 | 0.957–1.039 | 35% | −0.002 | +0.014 |
| 8 | POOL-KM-U | 0.082 [0.069, 0.112] | 0.918 | 0.858–0.984 | 100% | +0.026 | +0.047 |
| 8 | CONT-GA-PQPVC | 0.010 [0.005, 0.025] | 0.995 | 0.961–1.028 | 65% | +0.002 | +0.012 |

| $q$ | Method | E1 median [IQR] | $\hat{P}_f/P_{\mathrm{ref}}$ median | $\hat{P}_f/P_{\mathrm{ref}}$ range | Share $\hat{P}_f/P_{\mathrm{ref}} < 1$ | $\Delta\beta$ median | Worst $\Delta\beta$ |
|---|---|---|---|---|---|---|---|
| 8 | CONT-LBFGS | 0.042 [0.024, 0.060] | 0.958 | 0.928–1.023 | 95% | +0.013 | +0.023 |
| 8 | CONT-CMAES | 0.016 [0.006, 0.035] | 0.987 | 0.942–1.050 | 75% | +0.004 | +0.018 |
| 8 | CONT-SSO | 0.079 [0.050, 0.101] | 1.079 | 1.036–1.149 | 0% | −0.023 | −0.011 |
| 8 | SSOLB | 0.075 [0.038, 0.087] | 0.925 | 0.867–1.021 | 90% | +0.024 | +0.044 |

$\Delta\beta = \hat{\beta} - \beta_{\mathrm{ref}}$ (positive: reliability overstated). Worst $\Delta\beta$ is the largest $\Delta\beta$ over the 20 runs.

## 7. Discussion

### 7.1. Why the Hybrid Optimizer Did Not Help

Two properties of an inner optimizer must be kept apart. Optimizer quality is how close the selected point comes to the maximum of the learning function. Estimator quality is how much the selected point improves $\hat{P}_f$. The pre-specified comparisons measure the second; the first is only an intermediate step.

On optimizer quality, the L-BFGS-B polish did what it was designed to do. In the replayed states of the 23-bar truss (post-freeze addition), the polish improved the best SSO-stage value of criterion of Equation (4) in 25 of 25 states, and SSOLB, CMA-ES and CONT-LBFGS reached the best-found reference maximizer of Equation (4) (within 0.1 log units) in 88%, 96% and 84% of 25 states (Section 6.4; exploratory). The SSO stage spends 75% of $B$ on global exploration, which the density weight may make largely unnecessary on single-mode problems (a conjecture we did not test); the polish, not the swarm, produces the final point.

A better maximum is not a better point for estimating $P_f$. On the 23-bar truss the best-found reference maximizer of Equation (4) lay next to the design point $u^*$ (Section 6.4), where the density factor $\varphi_d(u)$ peaks along the limit state; we did not compute a reference maximizer on the other problems. In our runs, optimizers that maximized Equation (4) more completely tended to concentrate their points nearer $u^*$, with exceptions (CMA-ES on rackwitz_d100; Section 6.4). The rest of the failure domain can then stay false-safe, and $\hat{P}_f$ was biased low in the cases examined (Section 6.4; indicated by post-freeze analyses specified after Stage A11, with the limitations listed in Section 7.3). These analyses are consistent with the density factor as the cause but did not isolate it. Plain SSO, which maximized Equation (4) poorly, GA-PQPVC, which maximizes a different criterion, and the pool method AKMCS-U, which does not maximize Equation (4), spread their points along the limit state and were accurate at $q = 1$. The concentration is a property of where points are placed, not of continuous search as such: in a post hoc check, the pool batch method POOL-KM-U, whose points also lay close to $u^*$ on rackwitz_d10, also underestimated $P_f$ there. The density weight in Equation (4) may therefore play two roles. It may reduce the need for global exploration on single-mode problems, and may pull close maximizers towards $u^*$. The second effect is the opposite of what a more effective inner optimizer is meant to deliver.

A second, descriptive observation points the same way. On single-mode problems the pool methods select from $S$, which already covers the high-density part of the failure region in proportion to its probability. Continuous batches can instead cluster near existing design points (oscillator $q = 4$: SSOLB and CMA-ES placed points closer to existing design points and $\hat{\varepsilon}$ never reached 0.02; exploratory). In both cases a stronger maximizer of the learning function was not a better selector of points for $\hat{P}_f$.

**No cost advantage either.** Under the matched budget every continuous optimizer received the same number of learning-function evaluations per round, so SSOLB could gain only through better points, not through more search. The pre-specified cost comparison was dominated by censoring: in 47 of the 110 E2 cells both methods stopped at $N_{\max}$ in every seed (Section 6.1). Where the call counts did differ, at $d \geq 20$ on the multi-mode problems, the pre-specified H2 comparison favoured pool with Kriging-believer selection by 80–404 calls to the surrogate-based stopping criterion (series4_d20 and series4_d100, the latter a post-freeze addition), because POOL-KB-U reached that criterion in 17–18 of 20 runs and the best continuous methods in none, which is not a saving at equal, verified accuracy; the other pool batch methods had their median call counts at the cap where they were run (Section 6.1). The rule evaluates $\hat{\varepsilon}$ on the population $S$. A pool method places its points in $S$, whereas a continuous method places them anywhere in the box; this may explain part of the difference, but we did not test it. On inner-loop time (H3), SSOLB took about half the time per round of multistart L-BFGS-B. The pre-specified state-level tests favoured it in all 32 cells but ignore the dependence of states within a seed; the result is supported by the post hoc problem–seed analysis ($\rho < 1$ in all 55 units), measured on one GPU type and partly reflecting the implementation (CPU-based L-BFGS-B steps on a shared node; Section 6.1). This saving matters only when a limit-state call is cheap compared with a selection round of about 1–2.5 s; for an expensive $g$, $N_{\text{call}}$ dominates (Assumption A3).

**Dimension.** Whether the inner optimizer matters more at higher $d$ (H1) could not be answered in the pre-specified form, because $N_{\max}$ censored the call counts. On E1 (exploratory) the spread between the optimizers grew with $d$ on the Rackwitz family, mainly because plain SSO failed at $d = 100$ and SSOLB was far less accurate than the pool methods there (median E1 0.69 and 0.63 against about 0.012); on the multi-mode family it did not grow.

### 7.2. Safety Consequences of Mode Under-Learning

In every observed case the unlearned mode had lost failure mass and $\hat{P}_f$ was below $P_f$ (the learned modes did not compensate; Section 6.2 and Table 7). The error stopping rule can accept substantial underestimation, because $\hat{\varepsilon}$ is computed from the same surrogate: 17 of the 198 runs met $\hat{\varepsilon} \leq 0.02$ with $\hat{P}_f/P_f$ between 0.38 and 0.89 (11–62% below $P_f$; 13 of them with pool selection), and the other 181 exhausted their budget ($N_{\max}$) without meeting the criterion (Table 6). Multistart L-BFGS-B at $q = 1$ plausibly converges to the dominant design point from most restarts; plain SSO, GA-PQPVC and AKMCS-U placed points across the modes, whereas CMA-ES often did not on the two four-branch problems, and SSOLB left a mode unlearned in 15 of 20 four-branch runs (a mode never sampled in 8). The size of the bias depends on the share of the unlearned mode in $P_f$: from $\Delta\beta$ +0.025 at $d = 100$ (series4_d100, a post-freeze addition; small mode, 6.6% of $P_f$) to $\Delta\beta$ +0.835 in the S1 worst case. For a design check against a target $\beta$, the S1 worst case would accept a design whose reliability index is overstated by 0.8.

**Robustness of the finding.** Three observations bound it. First, the bias had one sign in every group of Table 6: all 198 runs with an unlearned mode had $\hat{P}_f$ below $P_f$ (an exploratory association, conditional on the estimated mode probabilities; S1 $q = 1$: 123; S1 $q = 4$ and 8: 53; Stage A10, on series4_d100: 22), whereas the runs with all modes learned were centred near 1 (medians 0.989–1.011). Second, it does not depend on the threshold for "learned": with [1/3, 3] instead of [0.5, 2] (Amendment A6 (b2)) all 14 Stage A10 decisions were unchanged, and in S1 only one unadjusted decision changed (four-branch, $q = 1$, SSOLB: $p = 0.016$ with [1/3, 3] against 0.062 with [0.5, 2]). The criterion itself was fixed after the S0 pre-tests of multistart L-BFGS-B, in which covered modes had $\hat{P}_k = 0$, and before S1 (Section 6.2). Third, it is not confined to continuous search: AKMCS-U stopped with a never-sampled mode in 3/20 four-branch runs (Section 6.3), and on series4_d100 one POOL-IF-U run ended with an unlearned mode and $\hat{P}_f/P_f = 3 \times 10^{-4}$.

**Consequence for practice.** Because $\hat{\varepsilon}$ can accept a run with an unlearned mode and a substantial underestimate (it accepted 17 of the 198 runs at $\hat{P}_f/P_f$ 0.38–0.89, 13 of them with pool selection; the other 181 exhausted their budget), a check must use information outside the surrogate's own error estimate: per-mode estimates $\hat{P}_k$ where the modes are known, or a coverage diagnostic of the selected points (Table 9). With $q = 1$ the selection method itself matters: at $d = 100$ (series4_d100), AKMCS-U, GA-PQPVC, plain SSO and SSOLB left no mode unlearned in 20 runs each, whereas multistart L-BFGS-B did so in 14 of 20.

### 7.3. Practical Selection Guide and Limitations

Table 9. When to use pool or continuous selection (evidence level in brackets).

| Situation | Recommended | Avoid | Evidence |
|---|---|---|---|
| Sequential selection ($q = 1$), any $d$ tested (2–100) | AKMCS-U (pool, $U$) | CONT-LBFGS on possibly multi-mode limit states; CMA-ES on four-branch-type problems; (descriptive) SSOLB on four-branch ($k = 7$) | H-MM (pre-specified); AKMCS-U vs continuous (exploratory) |
| Batch selection ($q = 4$, 8), single design point | POOL-KB-U or POOL-IF-U | SSOLB, plain SSO at $d = 100$ | H0-main (pre-specified); plain SSO (exploratory) |
| Batch selection, multi-mode, $d \leq 20$ | pool with Kriging-believer selection, POOL-KB-U (fewer calls to the stopping criterion on series4_d20), with a per-mode check; pool and continuous batches did not differ in mode learning | GA-PQPVC at $q = 8$ on four-branch | H2 (pre-specified, series4_d20); H-MM (pre-specified); unlearned-mode counts (descriptive) |
| Batch selection, multi-mode, $d = 100$ | pool with Kriging-believer selection, POOL-KB-U (fewest calls to the stopping criterion, descriptive; POOL-IF-U had its median at the call cap) | CMA-ES ($\hat{P}_f = 0$ cases) | H0-main E2 (pre-specified); H2 (pre-specified; on series4_d100, a post-freeze addition); safety (exploratory) |

| Situation | Recommended | Avoid | Evidence |
|---|---|---|---|
| Candidate pool infeasible (memory, very small $P_f$) | continuous search with a per-mode coverage check | — | not tested here (inference) |
| Any selection method (pool or continuous) | report per-mode $\hat{P}_k$ or a mode-coverage diagnostic | relying on $\hat{\varepsilon}$ alone (it can accept substantial underestimation) | Section 6.2 (exploratory; $\hat{\varepsilon}$ accepted 17 runs with an unlearned mode at $\hat{P}_f/P_f$ 0.38–0.89, 13 of them with pool selection) |
| Single design point, any selection method (continuous maximizer of Equation (4) or pool batch) | a check of point spread along the limit state; AKMCS-U at $q = 1$ | treating a better maximizer of Equation (4) as a better selector (CMA-ES, SSOLB at $q = 1$); selections concentrated near $u^*$ (for example POOL-KM-U on rackwitz_d10, post hoc) | Section 6.4 (indicated by follow-up analyses of existing runs) |

**Limitations.** The conclusions are bounded as follows.

- No results at $d = 40$ and no high-dimensional problem whose limit state is neither additive nor of low effective dimension (RP63 not run, Amendment A10).
- E2 is censored by $N_{\max}$ in 47 of the 110 H0-main cells, and the only equivalence under the TOST procedure of Amendment A9 (Stage A10; SSOLB and POOL-IF-U in calls on series4_d100) holds at the cap. H1 in its pre-specified ($N_{\text{call}}$) form is uninformative, and H2 is informative only on the two four-mode series-system problems (4 of the 6 pre-specified Block A cells); it is specific to pool with Kriging-believer selection (descriptively: POOL-KB-U and POOL-IF-U were not tested against each other), and some of its operational details were fixed at analysis time (Section 6.1).

- series4_d100 is a post-freeze addition: it was added after S1 because of the S1 H-MM result; the tests on it are pre-specified, the choice of the problem is not. It is series4_d20 with 98 instead of 18 dimensions that do not affect $g$, so it is high-dimensional for the surrogate and the search but not for the limit state.
- Two pre-specified methods (POOL-LP-U, CONT-PSO-R22) were not run (Amendment A13); a comparison with local-penalization batch selection [41] is left for future work.
- One GP model class, one stopping rule and one learning-function family per selection space; conclusions may differ with adaptive importance-sampling or subset-simulation pools [31–33] (truss56_orig used one fixed Gaussian proposal; Section 4.1).
- L-BFGS-B evaluations include gradients and count as one budget unit; pool methods are not budget-matched (by design).
- On the original 56-bar truss (truss56_orig) the candidate sets differ by design: the pool methods chose among importance-sampling points drawn around $u^*$, whereas the continuous methods maximized the $\varphi_d$-weighted criterion of Equation (4) over the box (Section 4.1). The median E1 of the U-minimizing pool methods AKMCS-U, POOL-KB-U and POOL-IF-U on this problem (0.002–0.004) is of the order of the sampling uncertainty of the population (standard deviation of $\hat{P}_f/P_f$ 0.0024; E1 with exact classification 0.0020 on average; Section 4.1), so the four SSOLB E1 losses against POOL-KB-U and POOL-IF-U on it may partly reflect this asymmetry (its other four losses there are against continuous optimizers with the same criterion); without the ten truss56_orig cells the H0-main E1 counts are 6/62/32 (Section 6.1).
- The inner budget $B$ was calibrated on the per-round use of unbudgeted multistart L-BFGS-B (Section 4.2), one of the compared optimizers. With 100 particles, plain SSO received about $B/100$ generations per round, from about 26 ($q = 1$ at $d = 20$ and 100) to about 173 (oscillator, $q = 8$), and about 88 for the 800-dimensional batch of series4_d100 at $q = 8$; SSOLB spends 75% of $B$ in SSO. The H0-main result is conditional on these budgets; the budget-sensitivity block (Block D: $4B$, one problem, five seeds) was inconclusive.
- The 56-bar truss and the 23-bar truss (a post-freeze addition) are linear in the loads given $E$ and $A$; neither is a costly nonlinear FE model.

- The engineering demonstration (23-bar truss) appears to have a single design point, so it tests accuracy and call count only; whether the mode under-learning of Section 6.2 occurs in an engineering structure with several failure modes is not tested here. All its runs stopped at $N_{\max}$, so it gives no call-count information.
- The concentration mechanism of Section 6.4 was found after the Stage A11 results were seen; the follow-up analyses that support it were specified afterwards, before their outputs were generated, and use existing runs (post-freeze additions outside the original families). The intervention adds one point per state and uses the same truss runs in which the concentration was observed; the cross-problem check is correlational and covers single-design-point problems only. In the intervention (distinct non-tied states), the first round was uninformative (3 of the 5 distinct round-0 states were tied; the other 2 favoured the pool choice), the pool choice was better in only 13 of 21 states from CONT-LBFGS runs (0.62) and in 10 of 14 from AKMCS-U runs (0.71), and the random $U < 2$ control beat the best-found reference maximizer only weakly (47 of 79 non-tied states, 0.595; one-sided sign test $p = 0.057$ over the states of the plan, in which this comparison was report-only; like the primary state-level test, it ignores the dependence of states within a seed and counts the identical round-0 states of a seed four times; the control was not tested at the seed level).
- H3 covers two optimizers and one state per run (the middle round), replayed on one GPU type; the time ratio includes implementation differences (CPU-based L-BFGS-B steps against GPU-based SSO) and the CPU load of a shared node (Amendment A12). All 55 replayed selections that were not bit-identical to the stored ones came from S1 states with $d \leq 20$, whose hyperparameters had been fitted from scratch in that round, whereas all 50 Stage A10 states ($d = 100$) were reproduced; since the GPU model and the software environment were the same, and states first run on the replay node also differed (6 of 20), the round-off-level differences most likely arise in rebuilding the GP from the saved hyperparameters, by a mechanism we did not identify.

## 8. Conclusions

1. With paired seeds, and with equal inner budgets for the continuous optimizers (the pool methods are not budget-matched), SSO with L-BFGS-B polishing did not improve accuracy or cost over gradient-based, evolutionary or pool-based selection. Against the three continuous optimizers that maximize the same criterion it won 6, tied 41 and lost 19 of 66 E1 comparisons, and against the two pool methods it won none, tied 23 and lost 21 of 44; over all 110 comparisons per metric it lost 40 E1 and 23 E2 comparisons and won 6 and 10, no E1 cell was equivalent under the TOST procedure of Amendment A9 (Stage A10), and its call counts were equivalent to POOL-IF-U only in two cells on series4_d100 (a post-freeze addition) in which both methods mostly stopped at the call cap. Its inner-loop time per round was about half that of multistart L-BFGS-B (H3): the pre-specified state-level tests favoured it in all 32 cells but ignore the dependence of states within a seed, and the result is supported by the post hoc problem–seed analysis ($\rho < 1$ in all 55 units); it was measured on one GPU type, partly reflects the implementation (CPU-based L-BFGS-B steps on a shared node) and matters only when limit-state calls are cheap.

2. The inner optimizer matters for safety: with sequential selection on series systems, multistart L-BFGS-B left a failure mode unlearned in most runs (pre-specified H-MM: 59 of 60 S1 runs at $d = 2$ and 20, and 14 of 20 at $d = 100$ on series4_d100; the learned-mode criterion was fixed after S0 pre-tests of multistart L-BFGS-B had shown covered modes with $\hat{P}_k = 0$, and before S1). In all 198 runs with an unlearned mode the learned modes did not compensate, and $\hat{P}_f$ was below $P_f$ (an exploratory association, conditional on the estimated mode probabilities; Tables 6 and 7; median $\hat{P}_f/P_f$ 0.93 and $\Delta\beta$ +0.025 at $d = 100$; worst S1 case $\hat{P}_f/P_f$ 0.06, $\Delta\beta$ +0.84). The error criterion can accept substantial underestimation: it accepted 17 of the 198 runs at $\hat{P}_f/P_f$ 0.38–0.89 (11–62% below $P_f$; 13 of them with pool selection), and the other 181 exhausted their budget.
3. A better optimizer of the learning function is not a better selector of points. Follow-up analyses of existing runs, specified after the initial observation and before their outputs were generated, indicate that, on single-design-point problems, selections concentrated near the design point leave the rest of the failure domain false-safe. On the 23-bar truss (a post-freeze addition), the optimizers that maximized the density-weighted criterion most completely (CMA-ES, SSOLB, CONT-LBFGS) produced such concentration, an observed tendency with exceptions (CMA-ES on rackwitz_d100), and so did a pool batch method on one problem (POOL-KM-U on rackwitz_d10, post hoc); the point is where selections land, not whether the search is continuous. For the three optimizers, $\hat{P}_f$ was biased low on the 23-bar truss ($q = 1$: median $\hat{P}_f/P_{\text{ref}}$ 0.31, 0.73 and 0.92). Adding the pool choice reduced the false-safe mass more than adding the best-found reference maximizer in all five seeds (one-sided seed-level sign test $p = 0.031$, the smallest one-sided value five seeds allow; 56 of 73 distinct non-tied states, descriptive because states share seeds), and across seven other single-design-point problems the distance of the selected points from the design point was positively rank-correlated with $\hat{P}_f/P_{\text{ref}}$ in 16 of 20 cells. These results have clear limits. The mechanism was found after the Stage A11 results were seen; the intervention adds only one point per state; the evidence covers single-design-point problems only, with negative correlations for the 56-bar truss at $q = 4$ and 8; the first round was uninformative (3 of 5 distinct states tied), and for states from CONT-LBFGS runs the pool choice was better in only 13 of 21 non-tied states; and the random $U < 2$ control beat the best-found reference maximizer only weakly (one-sided $p = 0.057$ at the state level, ignoring the dependence of states within a seed). Accuracy should therefore be judged on $\hat{P}_f$, not on how well the learning function is maximized.

4. Candidate-pool selection is the recommended default: AK-MCS at $q = 1$, and pool with Kriging-believer selection for batches. AK-MCS was never significantly less accurate than any continuous optimizer compared at $q = 1$ and never needed significantly more calls to the stopping criterion (exploratory). In the pre-specified H2 comparison (Block A, $q = 4$ and 8), the best pool method, chosen from the same results, satisfied the surrogate-based stopping criterion with 80–404 fewer calls than the best continuous method in all four informative cells, on the two four-mode series-system problems at $d = 20$ and $d = 100$; these are calls to the stopping criterion, not calls for an equal, verified accuracy. This result is specific to pool with Kriging-believer selection (POOL-KB-U): the other pool batch methods had their median call counts at the cap where they were run, as every continuous method did (POOL-KB-U and POOL-IF-U were not tested against each other), and a post hoc check with fixed pairs gave the same result. The two single-mode cells at $d = 100$ were censored, and the two $q = 1$ cells added at analysis time pointed the same way (exploratory). Any selection, continuous search in particular, should be combined with a per-mode or coverage diagnostic: the error criterion can accept substantial underestimation, and 13 of the 17 runs it accepted at $\hat{P}_f/P_f$ 0.38–0.89 used pool selection (Section 6.2).
5. Scope. The $d = 100$ evidence comes from two problems of simple structure: rackwitz_d100, whose limit state is a sum of independent terms, and series4_d100, whose limit state depends on 2 of its 100 variables. The 440 engineering follow-up runs on the linear-elastic 23-bar truss stopped at the call cap in every run. No claims are therefore made about high-dimensional problems whose limit state is neither additive nor of low effective dimension, or about expensive nonlinear models.

## Data and Code Availability

Available now to the journal's reviewers (not public): the supplementary material for review contains Part 1 of the code and data package, byte-identical to the file deposited in the Zenodo record (the code, configuration and container definitions, the seeds and run lists, the frozen plan with all amendments and their SHA-256 records, the problem definitions and reference values, the analysis outputs, the per-mode rebuild outputs and the figures), together with the plan of the follow-up analyses and the scripts and outputs of the post hoc checks (the robustness checks of H2 and H3, the direction audit, the re-evaluation of the saved GPs on an independent MC population, and the compensation, censoring, stop-reason, intervention and importance-sampling checks). Public on acceptance: the draft Zenodo record (DOI to be added) already holds, besides Part 1, the per-run JSON results (4,590 main runs: S1 4,010 and Stage A10 580; plus 440 engineering follow-up runs, Stage A11, a post-freeze addition) and the saved GP states (about 0.9 GB) used by the diagnostic and follow-up analyses. Still to be deposited before publication: the scripts and outputs of the post hoc checks named above, the plans of the Stage A11 diagnosis and of the follow-up analyses, the Stage A11 compute-cap decisions, the budget-calibration rules, the raw output of the H3 timing replay, the outputs of the budget pilots, of the 23-bar truss reference and verification computations, of the GP-predictor validation, of the CPU per-mode approximation and of the bit-identity checks, the commit log of the author's version-control repository and the cluster job records. Environment: Python 3.12, torch 2.14.0+cu130, BoTorch 0.18.1, GPyTorch 1.15.2, pycma 4.5.0, SciPy 1.18.1, NumPy 2.4.6; NVIDIA H200 GPUs (NCHC Nano4); seeds 0–19 (main), 1000–1004 (budget pilots).

## Author Contributions

Wei-Chang Yeh: Conceptualization, Methodology, Software, Validation, Formal analysis, Investigation, Data curation, Writing – original draft, Writing – review & editing, Visualization, Project administration, Funding acquisition.

**Acknowledgments**

The NVIDIA RTX PRO 6000 Blackwell Max-Q Workstation Edition GPU used for code development and for the validation checks of Section 4.1 was provided as in-kind hardware support through the NVIDIA Academic Grant Program. The author thanks the National Center for High-performance Computing (NCHC), Taiwan, for providing GPU resources on the Nano4 system under project MST115542.

**Funding**

This work was supported in part by the National Science and Technology Council (NSTC), Taiwan, R.O.C., under grant nos. NSTC 114-2221-E-007-123-MY3 and NSTC 113-2221-E-007-117-MY3.

**Declaration of Interests**

The author declares no known competing financial interests or personal relationships that could have appeared to influence the work reported in this paper.

**Declaration of Generative AI and AI-Assisted Technologies in the Writing Process**

During the preparation of this work the author used a generative AI assistant, Claude (Anthropic), used via Claude Code, to help draft and copy-edit text, to write and revise the experiment and analysis code, to prepare and submit the computing jobs, to run the analyses, to produce the figures and result tables, and to search and summarize the literature (Section 5.4). After using this tool, the author reviewed and edited the content as needed and takes full responsibility for the content of the publication.

**Appendix A. Additional tables**

Table S1. Amendments A1–A13 and the additions, procedures and analyses fixed after the freeze, in order of record (freeze: 2026-10-11, 13:41 UTC; Section 5.1 and Table 3).

| Item | Record, date and time (UTC); effective | Made after | Reason | Status in this paper |
| --- | --- | --- | --- | --- |
| Amendment A1: $\varepsilon$-stop may not trigger while $\hat{P}_f = 0$ | Amendment A1, 2026-10-11 13:47; S1 onward | an S0 pre-test showed the defect; before S1 | the stopping rule let runs stop with $\hat{P}_f = 0$ | routine amendment |
| Amendment A3: secondary hypothesis H-MM on missed modes (coverage, McNemar and full-coverage tests against CONT-LBFGS) | Amendment A3, 2026-10-12 00:30; 2026-10-12 | the freeze; before any S0 output | the frozen metrics did not measure missed failure modes | pre-specified (before any comparison output) |
| Amendment A2: platform moved to NCHC Nano4 H200; reproducibility gates | Amendment A2, 2026-10-12 00:31; 2026-10-12 | the freeze; before S0 outputs | — | routine amendment |

| Item | Record, date and time (UTC); effective | Made after | Reason | Status in this paper |
|---|---|---|---|---|
| Amendment A4: problems fourbranch_k6 and series4_d20 | Amendment A4, 2026-10-12 01:16; 2026-10-12 | the freeze; before any S0 result was viewed (one progress snapshot only) | H-MM needed more than one multi-mode problem; four-branch was the only one | pre-specified (before any comparison output) |
| Amendment A5: erratum, $B$ is calibrated per (problem, $q$) | Amendment A5, 2026-10-12 03:34; 2026-10-12 | the freeze; before S1 | erratum | routine amendment |
| Amendment A7: POOL-LP-U and CONT-PSO-R22 deferred | Amendment A7, 2026-10-12 12:17; 2026-10-13 | the freeze; before S1 | the full texts needed to verify their definitions were not available | routine amendment |
| Amendment A8: definitions of the Block C combinations | Amendment A8, 2026-10-12 12:17; 2026-10-13 | the freeze; before S1 | — | routine amendment |
| Amendment A6: learned-mode criterion (b), used in H-MM; sensitivity analyses (a), symmetric log error, and (b2), the [1/3, 3] variant of (b) | Amendment A6, 2026-10-12 22:21; 2026-10-13 | S0, which ran CONT-LBFGS only; before S1 | coverage does not show whether the probability of a mode is estimated | pre-specified (before any comparison output) |

| Item | Record, date and time (UTC); effective | Made after | Reason | Status in this paper |
|---|---|---|---|---|
| Amendment A9: TOST procedure for later stages; censored E2 cells | Amendment A9, 2026-10-14 02:52; 2026-10-14 | S1 | the frozen plan fixed the equivalence bounds but not the test procedure | procedure; S1 TOST exploratory |
| Amendment A10: scope after S1, rackwitz_d100 (from the frozen plan) and series4_d100 replace stages S1b and S2 | Amendment A10, 2026-10-14 03:37; 2026-10-14 | S1; before any Stage A10 run | the S1 H-MM result at $q = 1$ (population-based optimizers left fewer modes unlearned) | post-freeze addition (series4_d100) |
| Budget-calibration rule for series4_d100 (pilot runs that stopped by $\varepsilon$ with $\hat{P}_f = 0$ excluded in a check value; tolerance 5%, otherwise stop) | calibration rule, 2026-10-14 03:58 | S1; the stop reasons, rounds and $\hat{P}_f$ of 8 of the 15 pilot runs then finished; before any budget value was computed | the frozen pilot code (without Amendment A1) let runs stop with $\hat{P}_f = 0$ | procedure |

| Item | Record, date and time (UTC); effective | Made after | Reason | Status in this paper |
|---|---|---|---|---|
| Amendment A11: engineering demonstration, 23-bar truss (Stage A11), threshold 12 cm; outside all pre-specified families | Amendment A11, 2026-10-15 05:52; 2026-10-15 | S1 and Stage A10; before any Stage A11 run | a structural case with published reference values | post-freeze addition; exploratory |
| Budget-calibration rule and cost-projection stop rule for the 23-bar truss | calibration rule, 2026-10-15 05:57 | Amendment A11; before any Stage A11 reference, pilot or main result | the series4_d100 rule; the compute cap of Amendment A11 | procedure |
| Stage A11 compute cap raised from 8 to 9, 11 and 11.2 GPU-hours | author decisions, 2026-10-15 06:39, 10:43 and 13:05 | the cost projection of the Stage A11 pilot and of the running main stage | the pre-specified re-projection exceeded the current cap | procedure; methods, seeds, budgets, stopping and metrics unchanged |
| Diagnosis of the $q = 1$ underestimation on the 23-bar truss (verdict rules fixed in advance) | diagnosis plan, 2026-10-15 13:26 | the Stage A11 results (13:11) | CMA-ES, SSOLB and CONT-LBFGS underestimated $P_f$ at $q = 1$ | exploratory analysis of the Stage A11 runs; verdict "unresolved"; the concentration observation post hoc (Section 6.4) |

| Item | Record, date and time (UTC); effective | Made after | Reason | Status in this paper |
|---|---|---|---|---|
| Two follow-up analyses of the concentration mechanism, using existing runs | analysis plan, 2026-10-15 13:48 | Stage A11 and its diagnosis; before any of their outputs | to test the concentration observed post hoc on the 23-bar truss | post-freeze addition; outside the original families |
| Sensitivity of the cross-problem check (without plain SSO; without rackwitz_d100); POOL-KM-U observation on rackwitz_d10 | sensitivity script, 2026-10-15 23:43 (committed before it was run) | the result of the cross-problem check | requested by the author after that result was seen | post hoc, exploratory (Section 6.4) |
| Operational details of H1 and H2 (optimizer set, trend-test response, definition of "best"); $q = 1$ cells | analysis script, 2026-10-16 13:45 | S1 and Stage A10; before any H1 or H2 output | the frozen plan left them open | analysis choices; $q = 1$ cells exploratory |
| Amendment A12: H3 timing replay reduced (CONT-LBFGS vs SSOLB only, the middle-round state of each seed 0–4 run, three repetitions, one exclusive GPU, at most 8 GPU-hours) | Amendment A12, 2026-10-16 14:29; 2026-10-16 | S1 and Stage A10; before any replay timing | the replay of the frozen plan would have needed about 100 GPU-hours | procedure |

| Item | Record, date and time (UTC); effective | Made after | Reason | Status in this paper |
|---|---|---|---|---|
| Amendment A13: POOL-LP-U and CONT-PSO-R22 removed (pre-specified, not run) | Amendment A13, 2026-10-16 14:29; 2026-10-16 | all results; neither method had been run | the full texts needed to verify their definitions were not available before the experiments ended | routine amendment |
| H1 and H2 results restricted to the Block A cells | analysis script, 2026-10-17 04:20 | the H1 and H2 output (2026-10-16 13:49) and a review of the manuscript | the frozen plan placed H1 and H2 in Block A; the $q = 1$ cells had been added at analysis time | analysis choice |
| Robustness check of H2 with fixed pairs | analysis script, 2026-10-19 00:34 | the H2 result | an external pre-submission review | post hoc, exploratory |
| Split of the H0-main W/T/L counts into optimizer-only and criterion-changing cells; direction audit of the significant cells | analysis script, 2026-10-21 02:33 | all H0-main results | a second external pre-submission review | post hoc; no label changed (Sections 5.4 and 6.1) |
| Seed-level aggregation and seed-clustered bootstrap of H3 | analysis script, 2026-10-21 02:33 | the H3 result | a second external pre-submission review | post hoc, exploratory (Section 6.1) |

| Item | Record, date and time (UTC); effective | Made after | Reason | Status in this paper |
|---|---|---|---|---|
| Re-evaluation of the saved final GPs on an independent MC population: order of the methods by E1′; direction of the significant H0-main E1 cells | analysis script and GPU-hour estimate, 2026-10-21 02:33; ranking script 02:55; direction check 06:40 and gap 06:41, after the ranking output | all H0-main results | a second external pre-submission review | post hoc, exploratory (Section 6.1) |
| Compensation by the learned modes in the runs with an unlearned mode; partial censoring of the E2 cells; test details of H2; Holm-adjusted p of the single H0-main cells discussed in Section 6.1 | analysis script, 2026-10-21 08:38 | all H0-main, H2 and H-MM results | a third external pre-submission review | post hoc, exploratory (Sections 6.1 and 6.2); no label changed |

| Item | Record, date and time (UTC); effective | Made after | Reason | Status in this paper |
|---|---|---|---|---|
| Stop reasons of the mode-classified runs; H0-main W/T/L without truss56_orig; distinct intervention states and seed-level sign test; sampling uncertainty of the truss56_orig importance-sampling population; intervention fractions by round and source on the distinct states | analysis scripts, 2026-10-22 14:33 and 14:53 | all H0-main and H-MM results and the follow-up analyses | a fourth external pre-submission review and its targeted review | post hoc, exploratory (Sections 4.1, 6.1, 6.2, 6.4 and 7.3); no label changed |
| Other analyses not named in the frozen plan (for example the safety consequence of unlearned modes, Section 6.2, and the pool comparisons at $q = 1$, Section 6.3) | analysis scripts and reports after S1 and Stage A10 | the S1 or Stage A10 results | further analysis of the pre-specified results | exploratory, labelled where reported |

Record: the commit to the author's repository (UTC); effective: the date in Taiwan time (UTC+8), or the stage from which an amendment applies. —: no reason beyond the content. S0: pre-tests, inner-budget calibration and reproducibility checks (Section 5.3).

Table S2. Pre-specified H0-main comparisons, SSOLB − opponent (n = 20 paired seeds per cell; Holm over 110 tests per metric; Section 6.1 and Table 5).

| Problem | $q$ | Opponent | E1 median SSOLB / opp. | E1 HL [95% CI] | E1 $p_{\mathrm{H}}$ | E1 | E2 median SSOLB / opp. | E2 HL [95% CI] | E2 $p_{\mathrm{H}}$ | E2 |
|---|---|---|---|---|---|---|---|---|---|---|
| four-branch | 4 | CONT-LBFGS | 0.0148 / 0.0146 | 0.0022 [−0.0023, 0.0146] | 1 | T | 120 / 108 | 10.0 [4.0, 16.0] | 0.027 | L |
| four-branch | 4 | CONT-CMAES | 0.0148 / 0.125 | −0.0998 [−0.124, −0.077] | $2.1 \times 10^{-4}$ | W | 120 / 120 | 0.0 [−8.0, 0.0] | 1 | T |
| four-branch | 4 | CONT-SSO | 0.0148 / 0.0128 | 0.004 [−0.004, 0.0102] | 1 | T | 120 / 108 | 8.0 [2.0, 12.0] | 0.088 | T |
| four-branch | 4 | POOL-KB-U | 0.0148 / 0.016 | −0.0043 [−0.0283, 0.0046] | 1 | T | 120 / 112 | 6.0 [0.0, 12.0] | 1 | T |
| four-branch | 4 | POOL-IF-U | 0.0148 / 0.0154 | 0.0024 [−0.0066, 0.01] | 1 | T | 120 / 120 | 0.0 [−4.0, 0.0] | 1 | T |
| four-branch | 8 | CONT-LBFGS | 0.015 / 0.0158 | $-9.69 \times 10^{-17}$ [−0.0032, 0.0034] | 1 | T | 116 / 124 | −4.0 [−8.0, 0.0] | 1 | T |

| Problem | $q$ | Opponent | E1 median SSOLB / opp. | E1 HL [95% CI] | E1 $p_{\mathrm{H}}$ | E1 | E2 median SSOLB / opp. | E2 HL [95% CI] | E2 $p_{\mathrm{H}}$ | E2 |
|---|---|---|---|---|---|---|---|---|---|---|
| four-branch | 8 | CONT-CMAES | 0.015 / 0.0462 | −0.0296 [−0.042, −0.0158] | 0.026 | W | 116 / 124 | −8.0 [−16.0, −4.0] | 0.088 | T |
| four-branch | 8 | CONT-SSO | 0.015 / 0.0132 | $-8.00 \times 10^{-4}$ [−0.0053, 0.0014] | 1 | T | 116 / 120 | 0.0 [−4.0, 0.0] | 1 | T |
| four-branch | 8 | POOL-KB-U | 0.015 / 0.0144 | $-6.00 \times 10^{-4}$ [−0.0036, 0.0016] | 1 | T | 116 / 116 | 0.0 [−4.0, 8.0] | 1 | T |
| four-branch | 8 | POOL-IF-U | 0.015 / 0.0148 | $-8.00 \times 10^{-4}$ [−0.0078, 0.0053] | 1 | T | 116 / 124 | −8.0 [−16.0, −4.0] | 0.088 | T |
| fourbranch_k6 | 4 | CONT-LBFGS | 0.00834 / 0.0111 | −0.0016 [−0.00516, 0.00163] | 1 | T | 120 / 108 | 8.0 [4.0, 12.0] | 0.14 | T |
| fourbranch_k6 | 4 | CONT-CMAES | 0.00834 / 0.151 | −0.142 [−0.163, −0.121] | $2.1 \times 10^{-4}$ | W | 120 / 120 | −4.0 [−8.0, 0.0] | 0.35 | T |

| Problem | $q$ | Opponent | E1 median SSOLB / opp. | E1 HL [95% CI] | E1 $p_{\mathrm{H}}$ | E1 | E2 median SSOLB / opp. | E2 HL [95% CI] | E2 $p_{\mathrm{H}}$ | E2 |
|---|---|---|---|---|---|---|---|---|---|---|
| fourbranch_k6 | 4 | CONT-SSO | 0.00834 / 0.0105 | $-6.59 \times 10^{-4}$ [−0.00289, 0.00236] | 1 | T | 120 / 88 | 20.0 [15.0, 24.0] | 0.008 | L |
| fourbranch_k6 | 4 | POOL-KB-U | 0.00834 / 0.0102 | −0.00196 [−0.00606, 0.0018] | 1 | T | 120 / 112 | 6.0 [0.0, 16.0] | 1 | T |
| fourbranch_k6 | 4 | POOL-IF-U | 0.00834 / 0.0107 | $7.86 \times 10^{-4}$ [−0.00368, 0.0046] | 1 | T | 120 / 120 | −4.0 [−8.0, 0.0] | 0.35 | T |
| fourbranch_k6 | 8 | CONT-LBFGS | 0.012 / 0.0128 | 0.00146 [−0.00174, 0.00464] | 1 | T | 108 / 124 | −12.0 [−18.0, −8.0] | 0.038 | W |
| fourbranch_k6 | 8 | CONT-CMAES | 0.012 / 0.0511 | −0.0377 [−0.0521, −0.0231] | $7.8 \times 10^{-4}$ | W | 108 / 124 | −16.0 [−24.0, −12.0] | 0.014 | W |
| fourbranch_k6 | 8 | CONT-SSO | 0.012 / 0.0113 | $1.12 \times 10^{-4}$ [−0.00168, 0.00213] | 1 | T | 108 / 108 | 0.0 [−4.0, 4.0] | 1 | T |

| Problem | $q$ | Opponent | E1 median SSOLB / opp. | E1 HL [95% CI] | E1 $p_{\mathrm{H}}$ | E1 | E2 median SSOLB / opp. | E2 HL [95% CI] | E2 $p_{\mathrm{H}}$ | E2 |
|---|---|---|---|---|---|---|---|---|---|---|
| fourbranch_k6 | 8 | POOL-KB-U | 0.012 / 0.0123 | $2.66 \times 10^{-4}$ [−0.00258, 0.00236] | 1 | T | 108 / 100 | 4.0 [0.0, 8.0] | 1 | T |
| fourbranch_k6 | 8 | POOL-IF-U | 0.012 / 0.0123 | −0.00232 [−0.0117, 0.0034] | 1 | T | 108 / 124 | −16.0 [−24.0, −12.0] | 0.014 | W |
| oscillator | 4 | CONT-LBFGS | 0.0287 / 0.0057 | 0.0203 [0.0158, 0.0248] | $2.1 \times 10^{-4}$ | L | 160 / 98 | 62.0 [58.0, 66.0] | 0.008 | L |
| oscillator | 4 | CONT-CMAES | 0.0287 / 0.0243 | 0.0058 [0.00114, 0.00938] | 1 | T | 160 / 160 | 0.0 [0.0, 0.0] | 1 | Tc |
| oscillator | 4 | CONT-SSO | 0.0287 / 0.0109 | 0.0179 [0.0109, 0.0251] | 0.005 | L | 160 / 84 | 74.0 [70.0, 78.0] | 0.008 | L |
| oscillator | 4 | POOL-KB-U | 0.0287 / 0.00386 | 0.0228 [0.0161, 0.0294] | $3.4 \times 10^{-4}$ | L | 160 / 64 | 96.0 [94.0, 98.0] | 0.007 | L |

| Problem | $q$ | Opponent | E1 median SSOLB / opp. | E1 HL [95% CI] | E1 $p_{\mathrm{H}}$ | E1 | E2 median SSOLB / opp. | E2 HL [95% CI] | E2 $p_{\mathrm{H}}$ | E2 |
|---|---|---|---|---|---|---|---|---|---|---|
| oscillator | 4 | POOL-IF-U | 0.0287 / 0.0035 | 0.023 [0.017, 0.0296] | $3.4 \times 10^{-4}$ | L | 160 / 64 | 98.0 [96.0, 99.0] | 0.007 | L |
| oscillator | 8 | CONT-LBFGS | 0.00986 / 0.0056 | 0.0032 [-1.75 $\times 10^{-4}$, 0.0064] | 1 | T | 164 / 100 | 60.0 [56.0, 64.0] | 0.008 | L |
| oscillator | 8 | CONT-CMAES | 0.00986 / 0.00658 | 0.00344 [$2.80 \times 10^{-4}$, 0.00636] | 1 | T | 164 / 92 | 70.0 [64.0, 72.0] | 0.008 | L |
| oscillator | 8 | CONT-SSO | 0.00986 / 0.0143 | $-0.00515$ [$-0.00865$, $3.63 \times 10^{-5}$] | 1 | T | 164 / 92 | 68.0 [60.0, 76.0] | 0.008 | L |
| oscillator | 8 | POOL-KB-U | 0.00986 / 0.00496 | 0.00519 [0.00101, 0.00916] | 1 | T | 164 / 76 | 92.0 [88.0, 92.0] | 0.007 | L |
| oscillator | 8 | POOL-IF-U | 0.00986 / 0.0042 | 0.00477 [$6.32 \times 10^{-4}$, 0.0109] | 1 | T | 164 / 72 | 92.0 [88.0, 94.0] | 0.007 | L |

| Problem | $q$ | Opponent | E1 median SSOLB / opp. | E1 HL [95% CI] | E1 $p_{\mathrm{H}}$ | E1 | E2 median SSOLB / opp. | E2 HL [95% CI] | E2 $p_{\mathrm{H}}$ | E2 |
|---|---|---|---|---|---|---|---|---|---|---|
| Rackwitz $d = 10$ | 4 | CONT-LBFGS | 0.299 / 0.0724 | 0.21 [0.189, 0.239] | $2.1 \times 10^{-4}$ | L | 200 / 200 | 0.0 [0.0, 0.0] | 1 | Tc |
| Rackwitz $d = 10$ | 4 | CONT-CMAES | 0.299 / 0.019 | 0.274 [0.246, 0.295] | 0.006 | L | 200 / 200 | 0.0 [0.0, 0.0] | 1 | Tc |
| Rackwitz $d = 10$ | 4 | CONT-SSO | 0.299 / 0.0179 | 0.273 [0.244, 0.301] | $2.1 \times 10^{-4}$ | L | 200 / 200 | 0.0 [0.0, 0.0] | 1 | Tc |
| Rackwitz $d = 10$ | 4 | POOL-KB-U | 0.299 / 0.00752 | 0.282 [0.252, 0.309] | $2.1 \times 10^{-4}$ | L | 200 / 200 | 0.0 [0.0, 0.0] | 1 | Tc |
| Rackwitz $d = 10$ | 4 | POOL-IF-U | 0.299 / 0.0128 | 0.28 [0.248, 0.305] | $2.1 \times 10^{-4}$ | L | 200 / 200 | 0.0 [0.0, 0.0] | 1 | Tc |
| Rackwitz $d = 10$ | 8 | CONT-LBFGS | 0.0861 / 0.0522 | 0.0331 [0.0216, 0.0451] | 0.003 | L | 204 / 204 | 0.0 [0.0, 0.0] | 1 | Tc |
| Rackwitz $d = 10$ | 8 | CONT-CMAES | 0.0861 / 0.0118 | 0.0731 [0.0561, 0.0923] | $3.4 \times 10^{-4}$ | L | 204 / 204 | 0.0 [0.0, 0.0] | 1 | Tc |
| Rackwitz $d = 10$ | 8 | CONT-SSO | 0.0861 / 0.0696 | 0.0189 [0.00557, 0.033] | 1 | T | 204 / 204 | 0.0 [0.0, 0.0] | 1 | Tc |
| Rackwitz $d = 10$ | 8 | POOL-KB-U | 0.0861 / 0.00705 | 0.0747 [0.0601, 0.0921] | $2.1 \times 10^{-4}$ | L | 204 / 204 | 0.0 [0.0, 0.0] | 1 | Tc |

| Problem | $q$ | Opponent | E1 median SSOLB / opp. | E1 HL [95% CI] | E1 $p_{\mathrm{H}}$ | E1 | E2 median SSOLB / opp. | E2 HL [95% CI] | E2 $p_{\mathrm{H}}$ | E2 |
|---|---|---|---|---|---|---|---|---|---|---|
| Rackwitz $d = 10$ | 8 | POOL-IF-U | 0.0861 / 0.0118 | 0.0762 [0.0604, 0.0906] | $2.1 \times 10^{-4}$ | L | 204 / 204 | 0.0 [0.0, 0.0] | 1 | Tc |
| Rackwitz $d = 2$ | 4 | CONT-LBFGS | 0.0188 / 0.0158 | $-7.11 \times 10^{-4}$ [−0.00305, 0.00244] | 1 | T | 28 / 32 | −4.0 [−8.0, −2.0] | 0.35 | T |
| Rackwitz $d = 2$ | 4 | CONT-CMAES | 0.0188 / 0.0123 | 0.00505 [−0.00253, 0.0133] | 1 | T | 28 / 46 | −18.0 [−24.0, −12.0] | 0.009 | W |
| Rackwitz $d = 2$ | 4 | CONT-SSO | 0.0188 / 0.0182 | −0.00259 [−0.00376, $-1.02 \times 10^{-4}$] | 1 | T | 28 / 32 | −4.0 [−4.0, −2.0] | 0.032 | W |
| Rackwitz $d = 2$ | 4 | POOL-KB-U | 0.0188 / 0.0157 | 0.00325 [0.00142, 0.00508] | 0.16 | T | 28 / 32 | −4.0 [−4.0, −2.0] | 0.042 | W |
| Rackwitz $d = 2$ | 4 | POOL-IF-U | 0.0188 / 0.0149 | 0.00234 [$5.59 \times 10^{-4}$, 0.00427] | 0.38 | T | 28 / 32 | −4.0 [−4.0, −3.0] | 0.018 | W |

| Problem | $q$ | Opponent | E1 median SSOLB / opp. | E1 HL [95% CI] | E1 $p_{\mathrm{H}}$ | E1 | E2 median SSOLB / opp. | E2 HL [95% CI] | E2 $p_{\mathrm{H}}$ | E2 |
|---|---|---|---|---|---|---|---|---|---|---|
| Rackwitz $d = 2$ | 8 | CONT-LBFGS | 0.0186 / 0.00742 | 0.00804 [0.00204, 0.0134] | 1 | T | 36 / 44 | −12.0 [−16.0, −8.0] | 0.018 | W |
| Rackwitz $d = 2$ | 8 | CONT-CMAES | 0.0186 / 0.00996 | 0.00737 [0.00311, 0.0118] | 0.21 | T | 36 / 44 | −8.0 [−12.0, −4.0] | 0.024 | W |
| Rackwitz $d = 2$ | 8 | CONT-SSO | 0.0186 / 0.0162 | −0.00163 [−0.00335, -1.52 $\times 10^{-4}$] | 1 | T | 36 / 44 | −4.0 [−4.0, 0.0] | 0.33 | T |
| Rackwitz $d = 2$ | 8 | POOL-KB-U | 0.0186 / 0.0147 | 0.00564 [0.00224, 0.00772] | 0.28 | T | 36 / 44 | −4.0 [−8.0, −4.0] | 0.070 | T |
| Rackwitz $d = 2$ | 8 | POOL-IF-U | 0.0186 / 0.0118 | 0.00396 [0.00163, 0.0063] | 0.38 | T | 36 / 44 | −8.0 [−8.0, −4.0] | 0.010 | W |
| Rackwitz $d = 6$ | 4 | CONT-LBFGS | 0.0978 / 0.0164 | 0.0868 [0.0626, 0.107] | $4.7 \times 10^{-4}$ | L | 160 / 160 | 0.0 [0.0, 0.0] | 1 | Tc |

| Problem | $q$ | Opponent | E1 median SSOLB / opp. | E1 HL [95% CI] | E1 $p_{\mathrm{H}}$ | E1 | E2 median SSOLB / opp. | E2 HL [95% CI] | E2 $p_{\mathrm{H}}$ | E2 |
|---|---|---|---|---|---|---|---|---|---|---|
| Rackwitz $d = 6$ | 4 | CONT-CMAES | 0.0978 / 0.0875 | 0.00982 [−0.00226, 0.0212] | 1 | T | 160 / 160 | 0.0 [0.0, 0.0] | 1 | Tc |
| Rackwitz $d = 6$ | 4 | CONT-SSO | 0.0978 / 0.0219 | 0.075 [0.0525, 0.103] | $2.1 \times 10^{-4}$ | L | 160 / 160 | 0.0 [0.0, 0.0] | 1 | T |
| Rackwitz $d = 6$ | 4 | POOL-KB-U | 0.0978 / 0.0141 | 0.0901 [0.0689, 0.115] | $3.4 \times 10^{-4}$ | L | 160 / 136 | 24.0 [20.0, 28.0] | 0.008 | L |
| Rackwitz $d = 6$ | 4 | POOL-IF-U | 0.0978 / 0.0116 | 0.0911 [0.068, 0.117] | $3.4 \times 10^{-4}$ | L | 160 / 140 | 20.0 [16.0, 24.0] | 0.008 | L |
| Rackwitz $d = 6$ | 8 | CONT-LBFGS | 0.049 / 0.0126 | 0.0292 [0.0157, 0.0419] | 0.022 | L | 164 / 164 | 0.0 [0.0, 0.0] | 1 | Tc |
| Rackwitz $d = 6$ | 8 | CONT-CMAES | 0.049 / 0.0136 | 0.0316 [0.0194, 0.043] | 0.006 | L | 164 / 164 | 0.0 [0.0, 0.0] | 1 | Tc |
| Rackwitz $d = 6$ | 8 | CONT-SSO | 0.049 / 0.0429 | 0.00278 [−0.0131, 0.0161] | 1 | T | 164 / 164 | 0.0 [0.0, 0.0] | 1 | Tc |

| Problem | $q$ | Opponent | E1 median SSOLB / opp. | E1 HL [95% CI] | E1 $p_{\mathrm{H}}$ | E1 | E2 median SSOLB / opp. | E2 HL [95% CI] | E2 $p_{\mathrm{H}}$ | E2 |
|---|---|---|---|---|---|---|---|---|---|---|
| Rackwitz $d = 6$ | 8 | POOL-KB-U | 0.049 / 0.014 | 0.0315 [0.0197, 0.0464] | 0.007 | L | 164 / 148 | 20.0 [16.0, 24.0] | 0.009 | L |
| Rackwitz $d = 6$ | 8 | POOL-IF-U | 0.049 / 0.0132 | 0.0313 [0.0188, 0.0466] | 0.005 | L | 164 / 148 | 16.0 [12.0, 20.0] | 0.009 | L |
| series4_d20 | 4 | CONT-LBFGS | 0.0119 / 0.0131 | $2.85 \times 10^{-4}$ [−0.0198, 0.00328] | 1 | T | 301 / 301 | 0.0 [0.0, 0.0] | 1 | Tc |
| series4_d20 | 4 | CONT-CMAES | 0.0119 / 0.012 | 0.00107 [−0.00271, 0.00399] | 1 | T | 301 / 301 | 0.0 [0.0, 0.0] | 1 | T |
| series4_d20 | 4 | CONT-SSO | 0.0119 / 0.0125 | 0.00246 [-1.43 $\times 10^{-4}$, 0.00613] | 1 | T | 301 / 301 | 0.0 [0.0, 0.0] | 1 | Tc |
| series4_d20 | 4 | POOL-KB-U | 0.0119 / 0.0123 | $8.55 \times 10^{-4}$ [−0.0454, 0.00371] | 1 | T | 301 / 193 | 102.0 [66.0, 120.0] | 0.012 | L |
| series4_d20 | 4 | POOL-IF-U | 0.0119 / 0.0129 | $8.55 \times 10^{-4}$ [−0.0429, 0.00342] | 1 | T | 301 / 301 | 9.0 [0.0, 30.0] | 0.58 | T |

| Problem | $q$ | Opponent | E1 median SSOLB / opp. | E1 HL [95% CI] | E1 $p_H$ | E1 | E2 median SSOLB / opp. | E2 HL [95% CI] | E2 $p_H$ | E2 |
|---|---|---|---|---|---|---|---|---|---|---|
| series4_d20 | 8 | CONT-LBFGS | 0.0125 / 0.00998 | 0.0021 [−0.00172, 0.0124] | 1 | T | 301 / 301 | 0.0 [0.0, 0.0] | 1 | Tc |
| series4_d20 | 8 | CONT-CMAES | 0.0125 / 0.0146 | 0.00125 [−0.00314, 0.0132] | 1 | T | 301 / 301 | 0.0 [0.0, 0.0] | 1 | Tc |
| series4_d20 | 8 | CONT-SSO | 0.0125 / 0.0113 | 0.00214 [−0.00236, 0.0139] | 1 | T | 301 / 301 | 0.0 [0.0, 0.0] | 1 | Tc |
| series4_d20 | 8 | POOL-KB-U | 0.0125 / 0.0126 | $4.28 \times 10^{-4}$ [−0.00478, 0.0114] | 1 | T | 301 / 213 | 80.0 [56.0, 100.0] | 0.015 | L |
| series4_d20 | 8 | POOL-IF-U | 0.0125 / 0.0211 | −0.00374 [−0.0275, 0.00969] | 1 | T | 301 / 301 | 0.0 [0.0, 0.0] | 1 | T |
| 56-bar truss, mod. | 4 | CONT-LBFGS | 0.03 / 0.0314 | −0.0015 [−0.0038, $4.00 \times 10^{-4}$] | 1 | T | 212 / 212 | 0.0 [0.0, 0.0] | 1 | Tc |
| 56-bar truss, mod. | 4 | CONT-CMAES | 0.03 / 0.0248 | $6.00 \times 10^{-4}$ [−0.0042, 0.005] | 1 | T | 212 / 212 | 34.0 [0.0, 44.0] | 0.58 | T |

| Problem | $q$ | Opponent | E1 median SSOLB / opp. | E1 HL [95% CI] | E1 $p_{\mathrm{H}}$ | E1 | E2 median SSOLB / opp. | E2 HL [95% CI] | E2 $p_{\mathrm{H}}$ | E2 |
|---|---|---|---|---|---|---|---|---|---|---|
| 56-bar truss, mod. | 4 | CONT-SSO | 0.03 / 0.022 | 0.0106 [0.0052, 0.0148] | 0.25 | T | 212 / 206 | 18.0 [2.0, 32.0] | 0.088 | T |
| 56-bar truss, mod. | 4 | POOL-KB-U | 0.03 / 0.01 | 0.0204 [0.014, 0.0236] | 0.015 | L | 212 / 86 | 126.0 [122.0, 128.0] | 0.008 | L |
| 56-bar truss, mod. | 4 | POOL-IF-U | 0.03 / 0.0116 | 0.0205 [0.0122, 0.0228] | 0.15 | T | 212 / 104 | 110.0 [104.0, 114.0] | 0.008 | L |
| 56-bar truss, mod. | 8 | CONT-LBFGS | 0.0324 / 0.0292 | 0.0024 [$2.00 \times 10^{-4}$, 0.0048] | 1 | T | 212 / 212 | 0.0 [0.0, 0.0] | 1 | Tc |
| 56-bar truss, mod. | 8 | CONT-CMAES | 0.0324 / 0.016 | 0.018 [0.0118, 0.0222] | 0.092 | T | 212 / 212 | 0.0 [0.0, 0.0] | 1 | Tc |
| 56-bar truss, mod. | 8 | CONT-SSO | 0.0324 / 0.0178 | 0.0158 [0.0027, 0.0244] | 1 | T | 212 / 212 | 0.0 [0.0, 8.0] | 1 | T |
| 56-bar truss, mod. | 8 | POOL-KB-U | 0.0324 / 0.013 | 0.0238 [0.0205, 0.0264] | 0.16 | T | 212 / 92 | 120.0 [116.0, 124.0] | 0.008 | L |

| Problem | $q$ | Opponent | E1 median SSOLB / opp. | E1 HL [95% CI] | E1 $p_{\mathrm{H}}$ | E1 | E2 median SSOLB / opp. | E2 HL [95% CI] | E2 $p_{\mathrm{H}}$ | E2 |
|---|---|---|---|---|---|---|---|---|---|---|
| 56-bar truss, mod. | 8 | POOL-IF-U | 0.0324 / 0.011 | 0.026 [0.022, 0.0288] | 0.17 | T | 212 / 132 | 80.0 [76.0, 84.0] | 0.008 | L |
| 56-bar truss | 4 | CONT-LBFGS | 0.0225 / 0.0251 | −0.00312 [−0.00611, $-1.31 \times 10^{-4}$] | 1 | T | 212 / 212 | 0.0 [0.0, 0.0] | 1 | Tc |
| 56-bar truss | 4 | CONT-CMAES | 0.0225 / 0.0213 | $-6.30 \times 10^{-4}$ [−0.00378, 0.00426] | 1 | T | 212 / 212 | 0.0 [0.0, 0.0] | 1 | Tc |
| 56-bar truss | 4 | CONT-SSO | 0.0225 / 0.00492 | 0.0155 [0.0105, 0.0191] | 0.032 | L | 212 / 212 | 0.0 [0.0, 0.0] | 1 | Tc |
| 56-bar truss | 4 | POOL-KB-U | 0.0225 / 0.00202 | 0.0196 [0.016, 0.0233] | $2.1 \times 10^{-4}$ | L | 212 / 212 | 0.0 [0.0, 0.0] | 1 | Tc |
| 56-bar truss | 4 | POOL-IF-U | 0.0225 / 0.00281 | 0.0181 [0.0141, 0.0231] | $2.1 \times 10^{-4}$ | L | 212 / 212 | 0.0 [0.0, 16.0] | 1 | T |
| 56-bar truss | 8 | CONT-LBFGS | 0.028 / 0.0237 | 0.00551 [0.00349, 0.00769] | 0.015 | L | 212 / 212 | 0.0 [0.0, 0.0] | 1 | Tc |

| Problem | $q$ | Opponent | E1 median SSOLB / opp. | E1 HL [95% CI] | E1 $p_{\mathrm{H}}$ | E1 | E2 median SSOLB / opp. | E2 HL [95% CI] | E2 $p_{\mathrm{H}}$ | E2 |
|---|---|---|---|---|---|---|---|---|---|---|
| 56-bar truss | 8 | CONT-CMAES | 0.028 / 0.0194 | 0.0101 [0.00486, 0.0156] | 0.038 | L | 212 / 212 | 0.0 [0.0, 0.0] | 1 | Tc |
| 56-bar truss | 8 | CONT-SSO | 0.028 / 0.0108 | 0.0166 [0.0105, 0.0234] | 0.006 | L | 212 / 212 | 0.0 [0.0, 0.0] | 1 | Tc |
| 56-bar truss | 8 | POOL-KB-U | 0.028 / 0.00219 | 0.0264 [0.0238, 0.0289] | $2.1 \times 10^{-4}$ | L | 212 / 212 | 0.0 [0.0, 0.0] | 1 | T |
| 56-bar truss | 8 | POOL-IF-U | 0.028 / 0.00385 | 0.0247 [0.0223, 0.0273] | $2.1 \times 10^{-4}$ | L | 212 / 212 | 0.0 [0.0, 0.0] | 1 | T |
| Rackwitz $d = 100$ | 4 | CONT-LBFGS | 0.688 / 0.628 | 0.0581 [0.0266, 0.0676] | 0.17 | T | 1101 / 1101 | 0.0 [0.0, 0.0] | 1 | Tc |
| Rackwitz $d = 100$ | 4 | CONT-CMAES | 0.688 / 0.0738 | 0.6 [0.574, 0.623] | $2.1 \times 10^{-4}$ | L | 1101 / 1101 | 0.0 [0.0, 0.0] | 1 | Tc |
| Rackwitz $d = 100$ | 4 | CONT-SSO | 0.688 / 21.6 | −21.3 [−22.1, −20.4] | $2.1 \times 10^{-4}$ | W | 1101 / 1101 | 0.0 [0.0, 0.0] | 1 | Tc |
| Rackwitz $d = 100$ | 4 | POOL-KB-U | 0.688 / 0.0122 | 0.661 [0.619, 0.684] | 0.006 | L | 1101 / 1101 | 0.0 [0.0, 0.0] | 1 | Tc |
| Rackwitz $d = 100$ | 4 | POOL-IF-U | 0.688 / 0.0116 | 0.661 [0.619, 0.684] | $2.1 \times 10^{-4}$ | L | 1101 / 1101 | 0.0 [0.0, 0.0] | 1 | Tc |

| Problem | $q$ | Opponent | E1 median SSOLB / opp. | E1 HL [95% CI] | E1 $p_{\mathrm{H}}$ | E1 | E2 median SSOLB / opp. | E2 HL [95% CI] | E2 $p_{\mathrm{H}}$ | E2 |
|---|---|---|---|---|---|---|---|---|---|---|
| Rackwitz $d = 100$ | 8 | CONT-LBFGS | 0.628 / 0.494 | 0.129 [0.11, 0.155] | $2.1 \times 10^{-4}$ | L | 1101 / 1101 | 0.0 [0.0, 0.0] | 1 | Tc |
| Rackwitz $d = 100$ | 8 | CONT-CMAES | 0.628 / 0.0156 | 0.605 [0.583, 0.621] | $2.1 \times 10^{-4}$ | L | 1101 / 1101 | 0.0 [0.0, 0.0] | 1 | Tc |
| Rackwitz $d = 100$ | 8 | CONT-SSO | 0.628 / 31.1 | −30.7 [−31.8, −29.5] | $2.1 \times 10^{-4}$ | W | 1101 / 1101 | 0.0 [0.0, 0.0] | 1 | Tc |
| Rackwitz $d = 100$ | 8 | POOL-KB-U | 0.628 / 0.013 | 0.606 [0.588, 0.625] | $2.1 \times 10^{-4}$ | L | 1101 / 1101 | 0.0 [0.0, 0.0] | 1 | Tc |
| Rackwitz $d = 100$ | 8 | POOL-IF-U | 0.628 / 0.012 | 0.605 [0.589, 0.625] | $2.1 \times 10^{-4}$ | L | 1101 / 1101 | 0.0 [0.0, 0.0] | 1 | Tc |
| series4_d100 | 4 | CONT-LBFGS | 0.0416 / 0.0279 | 0.0138 [0.0067, 0.0217] | 0.11 | T | 1101 / 1101 | 0.0 [0.0, 0.0] | 1 | Tc |
| series4_d100 | 4 | CONT-CMAES | 0.0416 / 0.0123 | 0.0299 [0.0139, 0.0417] | 0.22 | T | 1101 / 1101 | 0.0 [0.0, 0.0] | 1 | Tc |
| series4_d100 | 4 | CONT-SSO | 0.0416 / 0.0111 | 0.0334 [0.0211, 0.0459] | 0.006 | L | 1101 / 1101 | 0.0 [0.0, 0.0] | 1 | Tc |

| Problem | $q$ | Opponent | E1 median SSOLB / opp. | E1 HL [95% CI] | E1 $p_{\mathrm{H}}$ | E1 | E2 median SSOLB / opp. | E2 HL [95% CI] | E2 $p_{\mathrm{H}}$ | E2 |
|---|---|---|---|---|---|---|---|---|---|---|
| series4_d100 | 4 | POOL-KB-U | 0.0416 / 0.0101 | 0.033 [0.0178, 0.0467] | 0.14 | T | 1101 / 677 | 404.0 [269.0, 454.0] | 0.010 | L |
| series4_d100 | 4 | POOL-IF-U | 0.0416 / 0.0105 | 0.0355 [0.0247, 0.0469] | 0.005 | L | 1101 / 1101 | 1.0 [0.0, 48.0] | 0.96 | T |
| series4_d100 | 8 | CONT-LBFGS | 0.0444 / 0.0406 | 0.0108 [0.00207, 0.0231] | 1 | T | 1101 / 1101 | 0.0 [0.0, 0.0] | 1 | Tc |
| series4_d100 | 8 | CONT-CMAES | 0.0444 / 0.0232 | 0.00146 [−0.441, 0.0363] | 1 | T | 1101 / 1101 | 0.0 [0.0, 0.0] | 1 | Tc |
| series4_d100 | 8 | CONT-SSO | 0.0444 / 0.0209 | 0.0221 [0.0136, 0.04] | 0.14 | T | 1101 / 1101 | 0.0 [0.0, 0.0] | 1 | Tc |
| series4_d100 | 8 | POOL-KB-U | 0.0444 / 0.0107 | 0.0396 [0.03, 0.0514] | 0.005 | L | 1101 / 713 | 350.0 [252.0, 424.0] | 0.010 | L |
| series4_d100 | 8 | POOL-IF-U | 0.0444 / 0.0105 | 0.0361 [0.0241, 0.0507] | 0.17 | T | 1101 / 1101 | 0.0 [0.0, 0.0] | 1 | T |

W/T/L: W if $p_{\mathrm{H}} < 0.05$ and the mean paired difference favours SSOLB (negative), L if it favours the opponent, otherwise T. c: censored E2 cell (both methods at $N_{\max}$ in all 20 seeds; Amendment A9). HL: Hodges–Lehmann estimate (median of the Walsh averages of the paired differences) with seed-bootstrap 95% CI ($10^4$ resamples). Totals: E1 W/T/L 6/64/40, E2 10/77/23, 47 censored E2 cells. Holm adjustment over the 110-test family (S1 90 and Stage A10 20 tests). The series4_d100 rows are a post-freeze addition (Section 5.1).

Table S3. Secondary hypothesis H-MM: the 38 tests significant in the 213-test Holm family (S1 171 and Stage A10 42 tests), each method against CONT-LBFGS (Section 6.2).

| Stage | Problem | $q$ | Method vs CONT-LBFGS | Test | Counts / HL | p | $p_H$ |
|---|---|---|---|---|---|---|---|
| A10 | series4_d100 | 1 | CONT-GA-PQPVC | unlearned mode, A6 (b) (McNemar) | 0/20 vs 14/20 unlearned | $1.2 \times 10^{-4}$ | 0.023 |
| A10 | series4_d100 | 1 | CONT-SSO | unlearned mode, A6 (b) (McNemar) | 0/20 vs 14/20 unlearned | $1.2 \times 10^{-4}$ | 0.023 |
| A10 | series4_d100 | 1 | SSOLB | unlearned mode, A6 (b) (McNemar) | 0/20 vs 14/20 unlearned | $1.2 \times 10^{-4}$ | 0.023 |
| A10 | series4_d100 | 1 | AKMCS-U | unlearned mode, A6 (b) (McNemar) | 0/20 vs 14/20 unlearned | $1.2 \times 10^{-4}$ | 0.023 |
| S1 | four-branch | 1 | CONT-GA-PQPVC | full-coverage $N_{call}$ (Wilcoxon) | HL −58.5 | $8.8 \times 10^{-5}$ | 0.017 |
| S1 | four-branch | 1 | CONT-SSO | full-coverage $N_{call}$ (Wilcoxon) | HL −59.5 | $9.5 \times 10^{-5}$ | 0.018 |
| S1 | four-branch | 1 | CONT-GA-PQPVC | missed mode (McNemar) | 0/20 vs 19/20 missed | $3.8 \times 10^{-6}$ | $8.0 \times 10^{-4}$ |
| S1 | four-branch | 1 | CONT-SSO | missed mode (McNemar) | 1/20 vs 19/20 missed | $7.6 \times 10^{-6}$ | 0.002 |
| S1 | four-branch | 1 | AKMCS-U | missed mode (McNemar) | 3/20 vs 19/20 missed | $3.1 \times 10^{-5}$ | 0.006 |

| Stage | Problem | $q$ | Method vs CONT-LBFGS | Test | Counts / HL | p | $p_{\mathrm{H}}$ |
|---|---|---|---|---|---|---|---|
| S1 | four-branch | 1 | CONT-GA-PQPVC | unlearned mode, A6 (b) (McNemar) | 2/20 vs 20/20 unlearned | $7.6 \times 10^{-6}$ | 0.002 |
| S1 | four-branch | 1 | CONT-SSO | unlearned mode, A6 (b) (McNemar) | 1/20 vs 20/20 unlearned | $3.8 \times 10^{-6}$ | $8.0 \times 10^{-4}$ |
| S1 | four-branch | 1 | AKMCS-U | unlearned mode, A6 (b) (McNemar) | 3/20 vs 20/20 unlearned | $1.5 \times 10^{-5}$ | 0.003 |
| S1 | fourbranch_k6 | 1 | CONT-GA-PQPVC | unlearned mode, A6 (b) (McNemar) | 0/20 vs 20/20 unlearned | $1.9 \times 10^{-6}$ | $4.1 \times 10^{-4}$ |
| S1 | fourbranch_k6 | 1 | CONT-SSO | unlearned mode, A6 (b) (McNemar) | 0/20 vs 20/20 unlearned | $1.9 \times 10^{-6}$ | $4.1 \times 10^{-4}$ |
| S1 | fourbranch_k6 | 1 | SSOLB | unlearned mode, A6 (b) (McNemar) | 1/20 vs 20/20 unlearned | $3.8 \times 10^{-6}$ | $8.0 \times 10^{-4}$ |
| S1 | fourbranch_k6 | 1 | AKMCS-U | unlearned mode, A6 (b) (McNemar) | 2/20 vs 20/20 unlearned | $7.6 \times 10^{-6}$ | 0.002 |
| S1 | series4_d20 | 1 | CONT-CMAES | full-coverage $N_{\mathrm{call}}$ (Wilcoxon) | HL −239.5 | $1.0 \times 10^{-4}$ | 0.020 |
| S1 | series4_d20 | 1 | CONT-GA-PQPVC | full-coverage $N_{\mathrm{call}}$ (Wilcoxon) | HL −236.8 | $1.0 \times 10^{-4}$ | 0.020 |

| Stage | Problem | $q$ | Method vs CONT-LBFGS | Test | Counts / HL | p | $p_H$ |
|---|---|---|---|---|---|---|---|
| S1 | series4_d20 | 1 | CONT-SSO | full-coverage $N_{call}$ (Wilcoxon) | HL −233.5 | $1.0 \times 10^{-4}$ | 0.020 |
| S1 | series4_d20 | 1 | SSOLB | full-coverage $N_{call}$ (Wilcoxon) | HL −223.8 | $1.0 \times 10^{-4}$ | 0.020 |
| S1 | series4_d20 | 1 | AKMCS-U | full-coverage $N_{call}$ (Wilcoxon) | HL −177.0 | $2.0 \times 10^{-4}$ | 0.036 |
| S1 | series4_d20 | 1 | CONT-CMAES | missed mode (McNemar) | 0/20 vs 14/20 missed | $1.2 \times 10^{-4}$ | 0.023 |
| S1 | series4_d20 | 1 | CONT-GA-PQPVC | missed mode (McNemar) | 0/20 vs 14/20 missed | $1.2 \times 10^{-4}$ | 0.023 |
| S1 | series4_d20 | 1 | CONT-SSO | missed mode (McNemar) | 0/20 vs 14/20 missed | $1.2 \times 10^{-4}$ | 0.023 |
| S1 | series4_d20 | 1 | SSOLB | missed mode (McNemar) | 0/20 vs 14/20 missed | $1.2 \times 10^{-4}$ | 0.023 |
| S1 | series4_d20 | 1 | AKMCS-U | missed mode (McNemar) | 1/20 vs 14/20 missed | $2.4 \times 10^{-4}$ | 0.043 |
| S1 | series4_d20 | 1 | CONT-CMAES | unlearned mode, A6 (b) (McNemar) | 1/20 vs 19/20 unlearned | $7.6 \times 10^{-6}$ | 0.002 |
| S1 | series4_d20 | 1 | CONT-GA-PQPVC | unlearned mode, A6 (b) (McNemar) | 0/20 vs 19/20 unlearned | $3.8 \times 10^{-6}$ | $8.0 \times 10^{-4}$ |
| S1 | series4_d20 | 1 | CONT-SSO | unlearned mode, A6 (b) (McNemar) | 0/20 vs 19/20 unlearned | $3.8 \times 10^{-6}$ | $8.0 \times 10^{-4}$ |
| S1 | series4_d20 | 1 | SSOLB | unlearned mode, A6 (b) (McNemar) | 0/20 vs 19/20 unlearned | $3.8 \times 10^{-6}$ | $8.0 \times 10^{-4}$ |

| Stage | Problem | $q$ | Method vs CONT-LBFGS | Test | Counts / HL | p | $p_{\mathrm{H}}$ |
|---|---|---|---|---|---|---|---|
| S1 | series4_d20 | 1 | AKMCS-U | unlearned mode, A6 (b) (McNemar) | 1/20 vs 19/20 unlearned | $7.6 \times 10^{-6}$ | 0.002 |
| S1 | four-branch | 4 | CONT-GA-PQPVC | full-coverage $N_{\mathrm{call}}$ (Wilcoxon) | HL +46.0 | $9.3 \times 10^{-5}$ | 0.018 |
| S1 | four-branch | 4 | POOL-KB-U | full-coverage $N_{\mathrm{call}}$ (Wilcoxon) | HL +36.0 | $1.4 \times 10^{-4}$ | 0.025 |
| S1 | fourbranch_k6 | 4 | CONT-GA-PQPVC | full-coverage $N_{\mathrm{call}}$ (Wilcoxon) | HL +16.0 | $1.5 \times 10^{-4}$ | 0.026 |
| S1 | series4_d20 | 4 | CONT-SSO | full-coverage $N_{\mathrm{call}}$ (Wilcoxon) | HL −26.0 | $2.7 \times 10^{-4}$ | 0.047 |
| S1 | four-branch | 8 | CONT-GA-PQPVC | full-coverage $N_{\mathrm{call}}$ (Wilcoxon) | HL +78.0 | $8.3 \times 10^{-5}$ | 0.016 |
| S1 | four-branch | 8 | CONT-GA-PQPVC | unlearned mode, A6 (b) (McNemar) | 17/20 vs 0/20 unlearned | $1.5 \times 10^{-5}$ | 0.003 |
| S1 | fourbranch_k6 | 8 | CONT-GA-PQPVC | full-coverage $N_{\mathrm{call}}$ (Wilcoxon) | HL +36.0 | $8.5 \times 10^{-5}$ | 0.017 |

McNemar: exact McNemar test on the paired seeds; Wilcoxon: paired Wilcoxon signed-rank test on the number of limit-state calls until all modes are covered, with the Hodges–Lehmann (HL) difference; p: unadjusted; $p_{\mathrm{H}}$: Holm-adjusted in the 213-test family. The Stage A10 tests are on series4_d100, a post-freeze addition (Section 5.1).